\documentclass[oneocolumn,secnumarabic,amssymb, nobibnotes, aps, prd]{revtex4-2}

\usepackage{cancel}
\usepackage{subcaption}
\newcommand{\dbar}{\mathchar '26 \mkern-10mu d}

\usepackage{physics}
\usepackage{graphicx}
\usepackage{float}
\usepackage{tabularx}
\usepackage{array}
\usepackage{booktabs}
\usepackage{amsmath}
\usepackage[T1]{fontenc}
\usepackage{hyperref}
\usepackage[spanish,english]{babel} % Configura español e inglés
\usepackage{natbib} % Para manejar las citas

\begin{document}

\title{Maximum power Stirling-like heat engine with a harmonically confined Brownian particle }

\author{Irene Prieto-Rodríguez $^{1}$, Antonio Prados $^{2}$ and Carlos A.~Plata $^{2,}$}
\email[corresponding author: ]{cplata1@us.es}
\affiliation{
$^1$ Department of Physics, Ludwig-Maximilians-Universität München, Schellingstr. 4, D-80799 Munich, Germany \\
$^2$ Física Teórica, Universidad de Sevilla, Apartado de Correos 1065, E-41080 Sevilla, Spain}
\date{December 2018}%

\begin{abstract}
Heat engines transform thermal energy into useful work, operating in a cyclic manner. For centuries, they have played a key role in industrial and technological development. Historically, only gases and liquids have been used as working substances, but the technical advances achieved over the past decades allow for expanding the experimental possibilities and designing engines operating with a single particle. In this case, the system of interest cannot be addressed at a macroscopic level and their study is framed in the field of stochastic thermodynamics.  In the present work, we study mesoscopic heat engines built with a Brownian particle submitted to harmonic confinement and immersed in a fluid acting as a thermal bath. We design a Stirling-like heat engine, composed of two isothermal and two isochoric branches, by controlling both the stiffness of the harmonic trap and the temperature of the bath. Specifically, we focus on the irreversible, non quasi-static, case---whose finite duration enables the engine to deliver a non-zero output power. This is a crucial aspect, which enables the optimisation of the thermodynamic cycle by maximising the delivered power---thereby addressing a key goal at the practical level. The optimal driving protocols are obtained by using both variational calculus and optimal control theory tools. Also, we numerically explore the dependence of the maximum output power and the corresponding efficiency on the system parameters. 
\end{abstract}
\maketitle

\section{Introduction}

As the size of a physical system is reduced, the importance of fluctuations grows, since they may become of the same order of magnitude as the meaningful average values. Present-day miniaturisation of technological devices has brought increasing attention to the extension of thermodynamic results to the mesoscale. Stochastic thermodynamics addresses this challenging goal, extending concepts such as heat, work and entropy to individual fluctuating trajectories governed by stochastic equations of motion, thereby incorporating time into the study of thermodynamic processes \cite{sekimoto_book,shiraishi_book,seifert_stochastic_2012}.  A colloidal particle immersed in a fluid at equilibrium  constitutes a paradigmatic system in this wide framework. Collisions of the colloidal particle with the fluid particles give rise both to a drag force and to thermal noise, related by the fluctuation-dissipation theorem~\cite{vankampen}. In addition, a confining force deriving from a potential may be added by means of optical traps \cite{ciliberto_experiments_2017}.

Leveraging the unavoidable fluctuations in a mesoscopic system in order to extract work, as Maxwell's demon would do \cite{noauthor_sorting_1879}, constitutes a revolutionary idea that has been deeply investigated---both theoretically and experimentally---in the last decades \cite{martinez_colloidal_2017}. On the one hand, ratchet models inspired by Feynman's work have motivated the design and study of Brownian motors \cite{Magnasco_ratchet_93,Reimann_motor_96,Julicher_motor_97,Reimann_motor_02,Hanggi_motor_05}. On the other hand, periodic driving of Brownian objects has allowed the transposition of classical thermodynamic cycles to the mesoscopic realm as stochastic heat engines  \cite{horowitz11,blickle_realization_2012,Holubec_logpot_14,Rana_Single_14,Tu_inertial_14,martinez15BrownianCarnotEngine,Bauer_limited_16,Dechant_underdamped_17,plata_building_2020,Tu_abstract_21}. Furthermore, other physical systems such as thermoelectric generators \cite{Apertet_thermoelectric_12,Apertet_thermoelectric2_12,Ouerdane_thermoelectric_15,Hua_thermoelectric_24},    two-level systems \cite{Gingrich_LargeDeviation_14} or active matter systems \cite{Krishnamurthy_active_16,Kumari_active_20} have been submitted to similar periodic drivings to build cyclic engines.

Performance of heat engines is usually measured in terms of power and efficiency. When driven in a quasi-static manner, as in the classical thermodynamic cycles, the output power vanishes. Thus, building mesoscopic finite-time counterparts of the classical heat engines (such as the Carnot, Stirling, Ericsson or Otto cycles) has become a relevant line of research. Renouncing reversibility entails a decrease in efficiency, i.e. there appears a trade-off between power and efficiency \cite{Shiraishi_trade-off_16,Pietzonka_trade-off_18}.
From a practical point of view, the problem of studying efficiency at maximum power has attracted many investigations in the field of finite-time thermodynamics. In this context, either general results in the low-dissipation or slow-driving regime \cite{curzon_efficiency_1975,van_den_broeck_eff,esposito_universality_2009,esposito_efficiency_2010,Wang_efficiency_12,tlili_finite_2012,Bauer_limited_16,Gonzalez-Ayala_trade-off_17,Frim_Geometric_bound_22,Frim_Optimal_22,Contreras_efficiency_23},
or specific results for a general driving in some models \cite{schmiedl_efficiency_2008,Rana_Single_14,Tu_inertial_14,Holubec_logpot_14,Dechant_underdamped_17,plata_building_2020,nakamura_fast_2020}  have been obtained. The former approach is based on a linear response theory scheme, in which the close-to-reversible protocols considered allow for expanding average heat and work around their reversible values, whereas the latter approach studies a concrete model for arbitrary driving, enabling the analysis of highly irreversible protocols beyond the linear response regime. 

The final objective of the present work is to design an optimal irreversible Stirling heat engine. The corresponding classical version encompasses four branches: two isothermal and two isochoric, all of them reversible---as already said, this entails that this classical engine delivers zero power. It is worth remarking that irreversible, but non-optimal, Stirling-like heat engines have been experimentally demonstrated with a harmonically confined colloidal particle \cite{blickle_realization_2012,martinez_colloidal_2017}. Here, we outperform this experimental realisation by analytically investigating how to optimise the cycle, specifically by maximising the delivered power. This is done by finding optimal driving protocols for the constituent processes, which has been a major topic in non-equilibrium statistical mechanics \cite{schmiedl_optimal_2007,plata_finite-time_2020,patron_thermal_2022}. The general framework of optimal control theory \cite{pontryagin,liberzon}, which facilitates the design of shortcuts and optimal connections between equilibrium states much faster than the corresponding natural relaxation, is extraordinarily useful in this context. In this regard, state-to-state transformations (SST) have been recently introduced to encompass  a rich set of techniques under the umbrella of control theory applied to statistical mechanics  \cite{martinez_engineered_2016,guery-odelin_driving_2023}.

The rest of the article is structured as follows. In section \ref{sec:the_model_system}, the dynamics of the model is introduced with special attention to the energetics of the processes. Section \ref{sec:building_blocks_for_a_Stirling_cycle} is devoted to the development of the constituent branches of the thermodynamic cycle, which are later employed to build the Stirling-like heat engine. General properties of the heat engine and a brief analysis of its reversible version are discussed in section \ref{sec:stirling_stochastic_heat_engine}. In section \ref{sec:opt}, the maximisation of the engine power is addressed. Therein, the corresponding efficiency at maximum power is also obtained. Finally, our conclusions are discussed in section \ref{ch:concl}.

%%%%

\section{The model system}\label{sec:the_model_system}

%%%%

\subsection{Harmonically confined Brownian particle}
We consider an overdamped Brownian particle immersed in a thermal bath at temperature $T$ and trapped in a one-dimensional harmonic potential with stiffness $k$. Both $T$ and $k$ can be externally controlled, i.e. we assume that their time-dependence can be tailored to our will. The friction coefficient is denoted by $\lambda$ and is considered as time-independent, even when the temperature $T$ depends on time. This may be surprising at first sight, since the friction coefficient for a colloidal particle is determined by both the particle geometry and the solvent viscosity---and the latter depends on the fluid temperature. Yet, it must be noted that the fluid temperature is usually not varied in experiments with optically trapped colloidal particles: the bath temperature is effectively raised by introducing external random forces, the amplitude of which can be controlled \cite{martinez_effective_13, martinez_colloidal_2017}.

Let $x$ denote the particle's position with respect to the centre of the trap. Its evolution follows the Langevin equation
\begin{equation}
	\lambda \frac{dx}{dt}(t)=-k(t)x(t)+\zeta(t),
	\label{langevin_eq}
\end{equation}
where $\zeta$ is a Gaussian white noise, 
\begin{equation}
	\langle \zeta(t)\rangle = 0,\quad 	\langle \zeta(t)\zeta(t')\rangle = 2\lambda k_BT(t)\delta(t-t').
	\label{white_noise_eqs}
\end{equation}
In the above equation, $k_B$ is the Boltzmann constant and $\delta$ is the Dirac delta function.

The linearity of Eq.~\eqref{langevin_eq} guarantees that, if the initial condition is distributed according to a Gaussian centred at $x=0$, then the distribution remains Gaussian, centred at the same point, at all times. Hence, the second moment $\langle x^2\rangle (t)$, i.e. the variance, fully determines the state of the system. This quantity evolves following the dynamic equation
\begin{equation}
	\lambda\frac{d}{dt}\langle x^2\rangle (t) = -2k(t)\langle x^2\rangle (t) + 2k_BT(t).\label{time_evol_variance}
\end{equation}

%%%%

\subsubsection{Energy, work and heat}

In the above system, we have a three-dimensional ``phase space'' $(k,\langle x^2 \rangle, T)$, a point in this space determines the instantaneous state of the system. The equilibrium equation of state is directly obtained by looking for the stationary solution of Eq.~\eqref{time_evol_variance}, when both $k$ and $T$ are time-independent:
\begin{equation}
 \langle x^2\rangle_{eq}=\frac{k_BT}{k}.
\label{equilibrium_cond}
\end{equation}
The energy of the system has a kinetic term and a configurational contribution stemming from the harmonic potential,
\begin{equation}
	E(t)=\frac12mv^2(t)+\frac12k(t)x^2,
\end{equation}
where $m$ is the mass of the Brownian particle and $v$ is its velocity. In the overdamped limit, which corresponds to our description, the latter variable is always at its equilibrium value: $\expval{v^2(t)}=k_BT(t)/m$. Therefore, the average energy is
\begin{equation}
	\expval{E(t)} = \frac{1}{2}k_BT(t)+\frac{1}{2}k(t)\expval{x^2(t)}.
	\label{energy}
\end{equation}
Thus, the equilibrium energy is characterised by the temperature of the system,
\begin{equation}
		\langle E\rangle _{eq}=k_BT.\label{eq_E}
\end{equation}
Taking differentials on both sides of Eq.~\eqref{energy}, one gets 
\begin{equation}
	d\expval{ E }=\underbrace{\frac{1}{2}k_B\,dT+\frac{1}{2}k\,d\expval{x^2 }}_{\dbar Q}+\underbrace{\frac12dk\,\expval{x^2}}_{\dbar  W}.
\end{equation}
On the one hand, we identify the infinitesimal average work $\;\dbar  W$ with the contribution to the energy variation
stemming from the change of the external mechanical parameters---namely, the stiffness of the trap $k$. On the other hand, we identify the infinitesimal average heat $\; \dbar  Q$  with the remainder of the energy variation, i.e. the energy variation stemming from the change of the probability distribution.  Note that $\; \dbar  Q$ includes two terms, corresponding to the change in the variance of the particle's velocity and position, respectively \cite{plata_finite-time_2020}. Since the velocity variable, as aforementioned, is always at equilibrium, the corresponding term is proportional to the temperature change. Although we have focused here on the average values of heat and work, the corresponding stochastic quantities can also be defined on individual fluctuating trajectories~\cite{sekimoto_book,shiraishi_book}.

Let us consider a process connecting two equilibrium state points: $(k_i,\langle x^2 \rangle_i, T_i)$ and $(k_f,\langle x^2 \rangle_f, T_f)$, where the subscripts $i$ and $f$ denote the initial and final situations, respectively. Work and heat associated with this transition are defined as follows,
\begin{align}
	&W_{i\to f}=\frac{1}{2}\int_i^fdk\,\langle x^2 \rangle,\\
	&Q_{i\to f} =\frac{k_B}{2}(T_f-T_i)+\frac12\int_i^fk\,d\langle x^2 \rangle.
\end{align}
Note that our sign convention considers energy transfers (both work and heat) from the environment to the system as positive, and as negative in the opposite case. Since our objective is to extract energy from our heat engine, we thus study cycles with a negative total work. The first law of thermodynamics is expressed as
\begin{equation}
	\Delta \langle E \rangle_{i\to f} = W_{i\to f}+Q_{i\to f},
\end{equation}
where $\Delta \langle E \rangle_{i\to f} = \langle E \rangle_f- \langle E \rangle_i$.

%%%%%

\subsection{Dimensionless variables}

In order to simplify our notation, let us introduce dimensionless variables for the physical properties that characterise our three-dimensional phase space. Specifically, we choose the units of $(k,\langle x^2 \rangle,T)$ to be normalised with respect to a reference equilibrium point $(k_{\text{ref}},\langle x^2 \rangle_{\text{ref}},T_{\text{ref}})$, i.e.
\begin{equation}
    \kappa \equiv \frac{k}{k_{\text{ref}}},\quad
y \equiv \frac{\langle x^2 \rangle}{\langle x^2 \rangle_{\text{ref}}}=\frac{k_{\text{ref}}}{k_BT_{\text{ref}}}\langle x^2 \rangle,\quad
\theta \equiv \frac{T}{T_{\text{ref}}}.\label{y_def}
\end{equation}
Note that the second equality for the normalised variance is a direct consequence of the equilibrium condition presented in Eq.~\eqref{equilibrium_cond}, which now reads
\begin{equation}
	\kappa y_{eq}=\theta.\label{dimensionless_equilibrium}
\end{equation}
Consistently, dimensionless average energy is defined,
\begin{equation}
	\mathcal{E}=\frac{\langle E \rangle}{k_BT_{\text{ref}}}.
\end{equation}
Thereupon, we can rewrite Eqs.~\eqref{energy} and \eqref{eq_E} in dimensionless form as
\begin{equation}
	\mathcal{E}=\frac12\theta+\frac12\kappa y, \quad \mathcal{E}_{eq}=\theta,
\end{equation}
where we have omitted the time-dependence in $\mathcal{E}$ to simplify the notation. Non-dimensional work and heat are defined consistently,
\begin{align}
	&\mathcal{W}_{i\to f}=\frac{1}{2}\int_i^fd\kappa\,y, \label{eq:def_W}\\
	&\mathcal{Q}_{i\to f} =\frac{1}{2}(\theta_f-\theta_i)+\frac12\int_i^fdy\,\kappa ,
\end{align}
and the first law now reads
\begin{equation}
	\Delta {\mathcal E}_{i\to f} =\mathcal W_{i\to f}+\mathcal Q_{i\to f}.
\end{equation}
In addition, non-dimensional time $\tau$ is defined as
\begin{equation}
	\tau \equiv t \frac{k_{\text{ref}}}{\lambda}.
\end{equation}
The state of the system at any time $\tau$ is thus characterised by the triplet $(\kappa,y,\theta)$. 
From now on, differentiation with respect to $\tau$ is denoted as $\dot{q}\equiv dq/d\tau$, for any physical quantity $q$. Hence, taking into account Eq.~\eqref{time_evol_variance}, the evolution of the non-dimensional variance is
\begin{equation}
	\dot{y}=-2\kappa y+2\theta.\label{variance_evol}
\end{equation}
The solution of Eq.~\eqref{variance_evol} can be written for general $\kappa$ and $\theta$ as
\begin{equation}
	y(\tau)=e^{-2\int_0^{\tau}d\tau\,\kappa(\tau')}\left[y_i+2\int_0^{\tau}d\tau'\,\theta(\tau')e^{2\int_0^{\tau'}d\tau''\,\kappa(\tau'')}\right].
\end{equation}
Note that, for the special case of constant stiffness $\kappa$ and temperature $\theta$, we have that the variance exponentially decays to its equilibrium value,
\begin{equation}
	y(\tau)=\frac{\theta}{\kappa}+e^{-2\kappa\tau}\left(y_i-\frac{\theta}{\kappa}\right). \label{eq:dyn-fixcont}
\end{equation}
Thus, for an infinitely slow---as compared with the characteristic relaxation time of our system---process, the stiffness of the trap and the temperature behave as constants in our timescale, the second term on the rhs vanishes, and the variance has its instantaneous equilibrium value for all times. This situation corresponds to the quasi-static limit, which is discussed in the upcoming sections.

%%%

\section{Building blocks for a Stirling cycle}\label{sec:building_blocks_for_a_Stirling_cycle}

The aim of this work is to build and investigate a non-equilibrium, irreversible, version of the Stirling cycle, in analogy to recent studies of irreversible Carnot-like heat engines \cite{schmiedl_efficiency_2008,martinez15BrownianCarnotEngine,plata_building_2020,nakamura_fast_2020}. 
Hence, our stochastic cycle encompasses isothermal and isochoric branches. We note that Stirling-like cycles have been recently explored in the linear response regime described in the introduction, specifically under the assumption of low dissipation \cite{Contreras_efficiency_23}. Herein, our analysis is performed for arbitrary driving, in general far from the low dissipation regime, for the specific model we are considering: a harmonically confined Brownian particle.

In the following, we define and analyse in detail the constituent branches of the Stirling-like engine.

\subsection{Isothermal processes}

In phase space, isothermal processes are represented by curves with a constant $\theta$ value. In this work, we consider two kinds of isothermal processes: quasi-static and optimal. Quasi-static processes are non-feasible experimentally and also lack practical interest, due to their null output power. Therefore, they are not our main focus but must be briefly analysed, since they are essential to study irreversible processes---as reference protocols for the latter. Our focus is put on irreversible isothermal processes, with a non-vanishing output power. In this regard, isothermal processes with maximum extracted work play a key role for our optimal Stirling-like heat engine.

In an isothermal process connecting equilibrium states, the average energy does not vary, since $\mathcal{E}_i=\mathcal{E}_f=\theta$. Therefore, by using the first law one concludes that
\begin{equation}\label{eq:first-law-isotherm}
\Delta\mathcal{E}_{i\to f}=0, \quad \widetilde{\mathcal{Q}}_{i\to f}=-\widetilde{\mathcal{W}}_{i\to f}
\end{equation}
for any isothermal process, either quasi-static or irreversible.

\subsubsection{Quasi-static isothermal processes}

A quasi-static process is defined as a succession of equilibrium states \cite{callen}. Therefore, an isothermal quasi-static process is represented by an equilibrium curve of the form given by Eq.~\eqref{dimensionless_equilibrium}, in which the temperature $\theta$ is fixed. To sweep this equilibrium curve $y(\tau)=\theta/\kappa(\tau)$, the control parameter $\kappa(\tau)$ must be varied sufficiently slowly.

For a quasi-static isothermal process, the work required to drive the system from the initial to the final state equals Helmholtz's free energy change $\Delta F_{i\to f}$, which is a state function. Then, work, heat and energy variation in such a process are
	\begin{equation}
		\mathcal{W}_{i\to f}^{QS} =\frac{\theta}2\ln\left(\frac{\kappa_f}{\kappa_i}\right)=\Delta F_{i\to f}=-\mathcal{Q}_{i\to f}^{QS}, \quad \Delta\mathcal{E}_{i\to f}^{QS}=0. \label{WqsIsoT_equals_DeltaF}
	\end{equation}

\subsubsection{Optimal isothermal processes}\label{subsect:optimal_isothermal}

Let us consider an isothermal process lasting a finite time $\tau_f$. In such a protocol, the system sweeps non-equilibrium states and thus work is a functional of the trajectory in phase space, i.e. a functional of the selected driving $\kappa(\tau)$. Therefore, it is meaningful to minimise the work performed on the system---i.e. maximise the work extraction---by looking for an optimal protocol $\widetilde{\kappa}(\tau)$. This optimisation problem has already been solved for both unconstrained \cite{schmiedl_efficiency_2008} and bounded stiffness \cite{plata_optimal_2019}. Note that we employ tilde throughout the paper for referring to the optimal protocols, the associated optimal paths in phase space, and the corresponding optimal values of the physical quantities.

The unconstrained optimal solution, which is obtained by solving the corresponding Euler-Lagrange equation, yields a linear evolution of $\sqrt{y}$ in time,
\begin{equation}
\widetilde{y}(\tau)=\left[\sqrt{y_i}+\left(\sqrt{y_f}-\sqrt{y_i}\right)\frac{\tau}{\tau_f}\right]^2,\quad \forall \tau\in\left[0,\tau_f\right].
\end{equation}
The optimal protocol $\widetilde{\kappa}$ stems from solving Eq.~\eqref{variance_evol} for the stiffness, 
\begin{equation}
	\widetilde{\kappa}(\tau)=\frac{\theta}{\widetilde{y}(\tau)}-\frac12\frac{d}{d\tau}\ln \widetilde{y}(\tau).
\end{equation}
Note that the $\widetilde{\kappa}$ is discontinuous at both the initial and final times, 
\begin{align}
	 \lim_{\tau\to0^+} \widetilde{\kappa}(\tau)&=\kappa_i-\frac{1}{\tau_f}\left(\sqrt{\frac{\kappa_i}{\kappa_f}}-1\right)\neq \kappa_i,\\
	 \lim_{\tau\to\tau_f^-} \widetilde{\kappa}(\tau)&=\kappa_f-\frac{1}{\tau_f}\left(1-\sqrt{\frac{\kappa_f}{\kappa_i}}\right)\neq \kappa_f.
\end{align}
 Similar discontinuities in the control parameter have repeatedly been found in stochastic thermodynamics \cite{plata_building_2020,patron_thermal_2022,guery-odelin_driving_2023,schmiedl_efficiency_2008,band82}. Note that the continuity at the boundaries is recovered in the quasi-static limit, in which $\tau_f\to\infty$.

The corresponding optimal work is obtained by substituting $\{\widetilde{\kappa},\widetilde{y}\}$ in  Eq.~\eqref{eq:def_W},
\begin{equation}
\widetilde{\mathcal{W}}_{i\to f}=\mathcal{W}_{i\to f}^{QS}+\frac{\theta}{\tau_f}\left(\frac{1}{\sqrt{\kappa_f}}-\frac{1}{\sqrt{\kappa_i}}\right)^2.\label{optimal_isothermal_work}
\end{equation}
Not only does the optimal work depend on the initial and final equilibrium states, but also on the process duration. The minimum irreversible work $\widetilde{\mathcal{W}}_{i\to f}-\mathcal{W}_{i\to f}^{QS}$ scales as $\tau_f^{-1}$, which vanishes in the quasi-static limit $\tau_f\to\infty$.
Albeit the extracted work is maximum in the quasi-static limit, it leads to a vanishing power output. The opposite limit, i.e. $\tau_f\to 0^+$, leads to the least energetically advantageous case: infinite work is required to perform such an instantaneous isothermal process.

\subsection{Isochoric processes}

In macroscopic heat engines operating with fluids, isochoric processes keep the volume constant and thus no work is done on---or extracted from---the system. In analogy with this situation, isochoric processes for a confined Brownian particle are defined as those in which the stiffness $\kappa$ is kept constant, since they also give zero work \cite{blickle_realization_2012}. Thus, isochoric processes are represented by curves with a fixed value for the trap stiffness $\kappa$ in phase space.

Isochoric processes are particularly simple: since work always vanishes, heat is fully determined by the temperature difference between the final and initial states, 
	\begin{equation}\label{eq:energy-balance-isochore}
		\mathcal{W}_{i\to f} =0, \quad
		\mathcal{Q}_{i\to f} = \Delta\mathcal{E}_{i\to f}
		=\theta_f-\theta_i.
	\end{equation}
        As one may intuitively expect, the system delivers heat to the bath in cooling processes, $\theta_f<\theta_i$, whereas it absorbs heat from the bath in heating processes, $\theta_f>\theta_i$.

We recall that our final goal is to build a maximum power Stirling-like heat engine. Therefore, given that work vanishes for any isochoric process, optimality here is associated with minimum connection time. In the following, we analyse quasi-static and optimal isochoric processes separately. 

\subsubsection{Quasi-static isochoric processes}

To sweep the equilibrium curve, the control parameter, which now is the bath temperature $\theta(\tau)$, must be varied infinitely slowly---i.e. the connection time $\tau_{f}$ must be very long. Aside from that, the energetics of such a quasi-static isochoric process is still given by Eq.~\eqref{eq:energy-balance-isochore}, since the reasoning leading thereto is independent of the duration of the process---i.e. independent on the intermediate states being equilibrium ones or not.

\subsubsection{Optimal isochoric processes}\label{subsect:optimal_isoch}

We now aim at studying the thermal optimal protocol that minimises the connection time between the equilibrium initial and final states, for fixed stiffness. This optimal shortcut---the thermal brachistochrone---has been investigated in depth in Ref.~\cite{patron_thermal_2022}. Therein, the problem was solved for arbitrary dimension, which yielded a rich phenomenology. Hereupon, we restrict ourselves to the one-dimensional case we are considering throughout.

The external control $\theta(\tau)$ is submitted to physical constraints, $\theta(\tau)\geq 0, \, \forall \tau$. Moreover, tighter bounds might be brought up by technical limitations in practice. Thus, we consider the general constraints $\theta_{\min}\leq\theta(\tau)\leq\theta_{\max}$ for the bath temperature, particularising later for the ideally relaxed conditions $\theta_{\min}\to0^+,\, \theta_{\max}\to\infty$.  The addressed optimisation problem with non-holonomic constraints cannot be solved with the tools of variational calculus. Instead, less restrictive methods of optimal control theory, like Pontryagin's maximum principle, are needed \cite{pontryagin,liberzon}.

The minimum time isochoric connection is a bang-bang protocol without switchings---i.e. a protocol in which $\theta(\tau)$ is equal to one of its bounds for the whole duration or the process. Specifically, the optimal control $\widetilde{\theta}(\tau)$ is \cite{patron_thermal_2022}
\begin{equation}
  \widetilde{\theta}(\tau)=
  \widetilde{\theta}\equiv\begin{cases}\theta_{\max},&\text{if}\quad \theta_i<\theta_f,\\\theta_{\min},&\text{if}\quad \theta_i>\theta_f,
	\end{cases}\quad\quad\quad\quad \forall\tau\in\left(0,\widetilde{\tau_f}\right).\label{opt_theta}
\end{equation}
 Again, the optimal control presents finite-jumps at the initial and final times: $\widetilde{\theta}\neq \theta_{i,f}$.
 
Since the optimal control is constant in the open interval $\left(0,\widetilde{\tau_f}\right)$, the evolution of the variance is simply given by Eq.~\eqref{eq:dyn-fixcont}. Enforcing that $y_i$ must evolve up to $y_f$ in a time $\widetilde{\tau_f}$, the latter is found to be
\begin{equation}
  \widetilde{\tau}_f=\begin{cases}\dfrac{1}{2\kappa}\ln\left(\dfrac{\theta_{\max}-
        \theta_i}{\theta_{\max}-\theta_f}\right),&\text{if}\quad \theta_i<\theta_f,\\[3ex]
\dfrac{1}{2\kappa}\ln\left(\dfrac{\theta_i-\theta_{\min}}{\theta_f-\theta_{\min}}\right),&\text{if}\quad \theta_i>\theta_f.
	\end{cases}
	\label{eq:tmins}
\end{equation}
It is worth highlighting that, although the control parameter $\widetilde{\theta}(\tau)$ is not continuous at both ends of the isochoric process, the associated optimal variance $\widetilde{y}$ is continuous in the whole time interval.

The energetics of the above optimal isochoric process is again given by Eq.~\eqref{eq:energy-balance-isochore}, it equals the quasi-static one---recall that work identically vanishes and thus heat equals the energy variation. As previously mentioned, the key difference between quasi-static and optimal isochoric processes is the latter's lasting a finite time.

In our discussion, we are particularly interested in the ideal situation $\theta_{\min}\to 0^+,\, \theta_{\max}\to\infty$, for which Eq.~\eqref{eq:tmins} simplifies to
\begin{equation}
	\widetilde{\tau}_f=\begin{cases}0,&\text{if}\quad \theta_i<\theta_f,\\
	\dfrac{1}{2\kappa}\ln\left(\dfrac{\theta_i}{\theta_f}\right),&\text{if}\quad \theta_i>\theta_f.
	\end{cases}\label{optimal_time_brachistochrone}
      \end{equation}
In the ideal limit of infinite heating power, our relaxing of the upper bound $\theta_{\max}\to +\infty$ leads to an optimal instantaneous connection for heating processes, whereas the physical lower bound $\theta_{\min}\to 0^{+}$ leads to a non-instantaneous optimal connection for cooling processes.

%%%%

\section{Stochastic Stirling heat engine}\label{sec:stirling_stochastic_heat_engine}

In this section, we study the stochastic version of a Stirling cycle performed by our model system. Experimental versions of such engines have been experimentally built in the last decades \cite{blickle_realization_2012,martinez_colloidal_2017}.\\

\begin{figure}%[h!]
	\centering
	\captionsetup{width=.85\linewidth}
	\includegraphics[width=1 \textwidth]{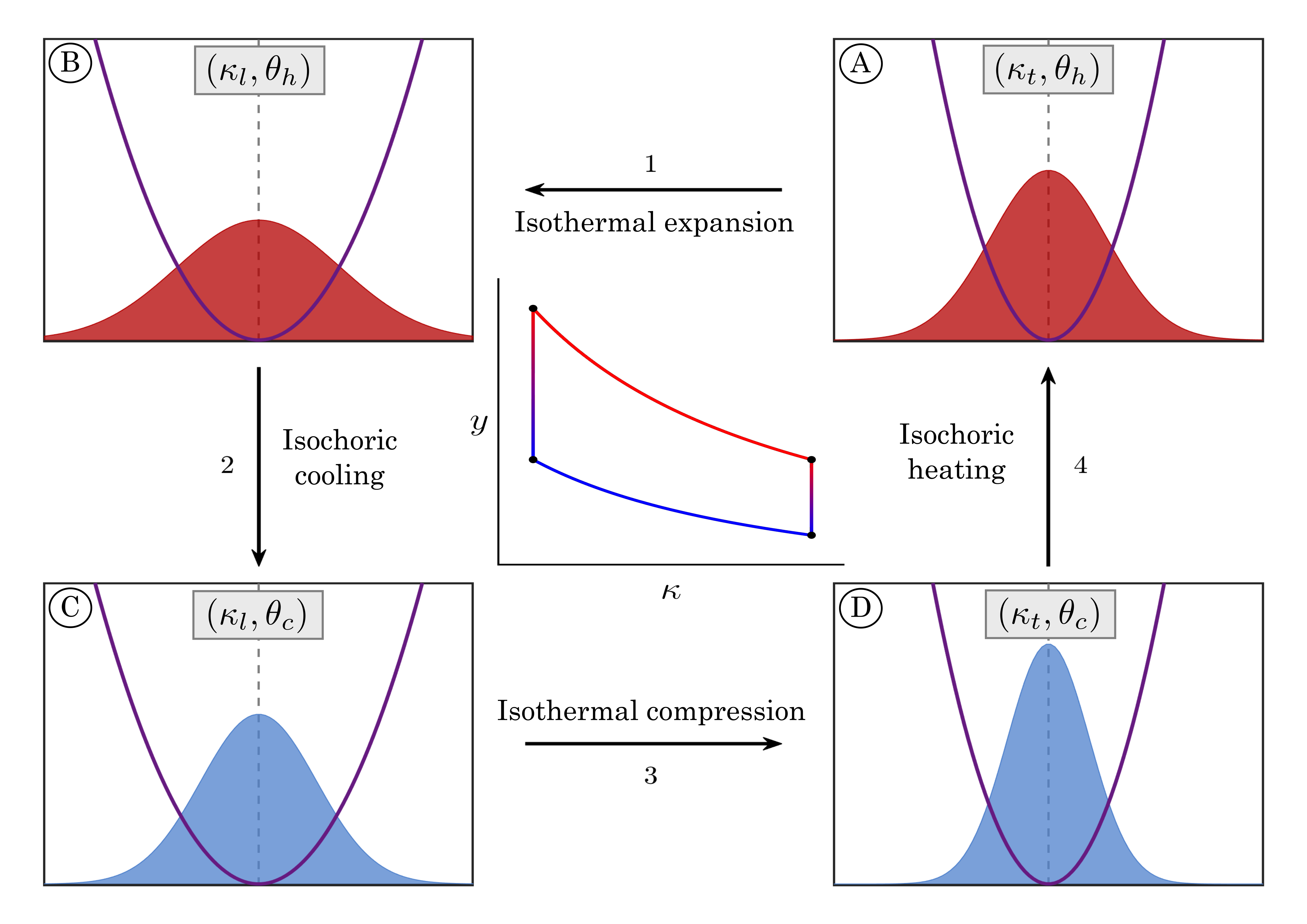}
	\caption{Scheme of the stochastic Stirling cycle. The harmonic confining potential at the operating points of the cycle is represented by the purple curves. The filled red and blue areas correspond to the probability density functions at those state-points, where red (blue) refers to the hot (cold) equilibrium temperatures. The representation of the heat engine in the $(\kappa,y)$ plane corresponds to the quasi-static version of the described cycle.}
	\label{fig:esquemaStirling}
\end{figure}

Analogously to the classical Stirling heat engine, our cyclic process encompasses four strokes, which are illustrated in Fig.~\ref{fig:esquemaStirling}:
\begin{enumerate}
\item \textbf{Isothermal expansion} at the hot bath temperature $\theta_h\equiv \theta_A=\theta_B$, connecting the phase points $A$ and $B$, i.e. $\left(\kappa_A,y_A,\theta_A\right)$ and $\left(\kappa_B,y_B,\theta_B\right)$. The confining harmonic potential is modified by controlling its stiffness $\kappa(\tau)$; the term \textit{expansion} (\textit{compression}) refers to the sign of the stiffness increment, i.e. to $\Delta k<0$ ($\Delta k>0$).
	\item \textbf{Isochoric 
          cooling} at \textit{loose} stiffness $\kappa_l\equiv\kappa_B=\kappa_C$, starting from the state-point $B\equiv\left(\kappa_B,y_B,\theta_B\right)$ up to $C\equiv\left(\kappa_C,y_C,\theta_C\right)$, with $\theta_{C}<\theta_{B}$. The time-dependence of the temperature of the heat bath $\theta(\tau)$ is now controlled.
	\item \textbf{Isothermal compression} at the cold bath temperature $\theta_c\equiv\theta_C=\theta_D<\theta_h$, linking states $C\equiv\left(\kappa_C,y_C,\theta_C\right)$ and $D\equiv\left(\kappa_D,y_D,\theta_D\right)$. As in process 1, the control variable is the time-dependent stiffness of the harmonic trap.
	\item \textbf{Isochoric 
          heating} at \textit{tight} stiffness $\kappa_t\equiv\kappa_D=\kappa_A>\kappa_l$, departing from state $D\equiv\left(\kappa_D,y_D,\theta_D\right)$ and closing the cycle by returning to the initial point $A\equiv\left(\kappa_A,y_A,\theta_A\right)$; thus $\theta_{A}>\theta_{D}$. As in process 2, the control variable is the time-dependent temperature of the bath.
\end{enumerate}

On the one hand, in the isotherms 1 and 3, the absorbed (released) heat equals the delivered (consumed) work,
\begin{equation}\label{eq:W-Q-AB-CD}
	\mathcal{W}_{AB}=-\mathcal{Q}_{AB}<0,\qquad 	\mathcal{W}_{CD}=-\mathcal{Q}_{CD}>0.
\end{equation}
as given by Eq.~\eqref{eq:first-law-isotherm}. On the other hand, in the isochores 2 and 4 of our cycle,
\begin{align}
	&\mathcal{W}_{BC}=0, &\mathcal{Q}_{BC}=\theta_C-\theta_B=\theta_c-\theta_h<0; \label{eq:QBC}\\
	&\mathcal{W}_{DA}=0, &\mathcal{Q}_{DA}=\theta_A-\theta_D=\theta_h-\theta_c>0 \label{eq:QDA}.
\end{align}
as given by Eq.~\eqref{eq:energy-balance-isochore}.

We are interested in building a heat engine, and thus we want our device to convert heat extracted from the baths to work. The absorbed heat corresponds to the first isothermal branch $A\to B$: $\mathcal{Q}_{AB}$. The total work in the Stirling-like cycle is
\begin{equation}
	\mathcal{W}\equiv\mathcal{W}_{AB}+\cancelto{0}{\mathcal{W}}_{BC}+\mathcal{W}_{CD}+\cancelto{0}{\mathcal{W}}_{DA}=\mathcal{W}_{AB}+\mathcal{W}_{CD}.
\end{equation}
The efficiency of our stochastic heat machine is defined, in analogy with macroscopic thermodynamics, as the ratio of the performed work over the extracted heat,
\begin{equation}\label{eq:eta-func-W-Q}
\eta\equiv \frac{-\mathcal{W}}{\mathcal{Q}_{AB}}=\frac{-\left(\mathcal{W}_{AB}+\mathcal{W}_{CD}\right)}{\mathcal{Q}_{AB}}=1-\frac{\mathcal{W}_{CD}}{\mathcal{Q}_{AB}}<1,
\end{equation}
where we have brought to bear Eq.~\eqref{eq:W-Q-AB-CD}.

It must be remarked that we have not included the absorbed heat in the isochore $D \to A$ in our definition of the efficiency. As expressed by Eqs.~\eqref{eq:QBC} and \eqref{eq:QDA}, $\mathcal{Q}_{BC}$ and $\mathcal{Q}_{DA}$ are equal in absolute value but differ in sign. Therefore, a regeneration mechanism may recycle the heat yielded in the isochore $B \to C$ for being subsequently absorbed in the isochore $D \to A$ \cite{Contreras_efficiency_23}. The inclusion of this regeneration mechanism increases the efficiency, making it possible to reach Carnot's limit in the quasi-static regime, as shown in Sec. \ref{sec:quasiS}. The use of a regenerator is common in Stirling and Ericsson cycles \cite{finite_thermo_book}. In classical designs with working fluids, the regenerator serves as a temporary thermal energy storage device that absorbs heat during one part of the cycle and later transfers it back to the working fluid.

Let us denote the time duration of each branch by $\tau_{AB}$, $\tau_{BC}$, $\tau_{CD}$, $\tau_{DA}$, respectively. Thence, the delivered power in the cycle is
\begin{equation}
	\mathcal{P}\equiv \frac{-\mathcal{W}}{\tau_{AB}+\tau_{BC}+\tau_{CD}+\tau_{DA}}=\frac{-\left(\mathcal{W}_{AB}+\mathcal{W}_{CD}\right)}{\tau_{AB}+\tau_{BC}+\tau_{CD}+\tau_{DA}}.\label{power}
\end{equation}
If we temporarily forgot about the constraints to which our engine is submitted, we would expect that $4\times3=12$ parameters were necessary to fully depict the cycle characterised by 4 points in a 3-dimensional phase space. Nonetheless, we now consider normalisation on the phase space coordinates with respect to the initial state, which is equivalent to take $k_{\text{ref}}=k_A$ and $T_{\text{ref}}=T_A$ in Eq.~\eqref{y_def}, and thence point $A$ is fixed: $\left(\kappa_A,y_A,\theta_A\right)=(1,1,1)$. Furthermore, the operating points describe equilibrium states and thus the corresponding condition, given in Eq.~\eqref{dimensionless_equilibrium}, imposes three additional constraints,
\begin{equation}
	\kappa_B y_B=\theta_B,\quad	\kappa_C y_C=\theta_C,\quad	\kappa_D y_D=\theta_D .
\end{equation}
Moreover, as a consequence of the branches being isothermal and isochoric, two more pairs of restrictions are added,
\begin{align}
\theta_A&=\theta_B,\qquad\theta_C=\theta_D,\\
\kappa_B&=\kappa_C,\qquad\kappa_A=\kappa_D.
\end{align}

The above discussion entails that the operating points of the Stirling cycle are uniquely defined by two parameters. We characterise thus the cycle by (i) the temperature ratio
\begin{equation}
	\nu \equiv \frac{\theta_c}{\theta_h}=\theta_c<1,
\end{equation} 
and (ii) the compression ratio
\begin{equation}
	\chi\equiv\frac{\kappa_l}{\kappa_s}=\kappa_l<1.
\end{equation}
The phase coordinates of the operating points of the cycle as a function of the chosen variables $(\nu,\chi)$ are collected in Table \ref{table:operating_points}. Our choice of dimensionless units and parameters to describe the Stirling-like heat engine make it possible to compare its performance with that of the Carnot-like heat engine analysed in Ref.~\cite{plata_building_2020} directly in dimensionless variables, since the dimensionless units to describe the two cycles are the same.
\begin{table}[H] 
\caption{Operating points of the stochastic Stirling heat engine. See also Fig.~\ref{fig:proyeccioncicloStirling}.
	\label{table:operating_points}}
\newcolumntype{C}{>{\centering\arraybackslash}X}
\begin{tabularx}{\textwidth}{CCCC}
\toprule
\,	& $\kappa$	& $y$ & $\theta$ \\
\midrule
\end{tabularx}
\begin{tabularx}{\textwidth}{C|CCC}
$A$		& $1$			& $1$	& $1$\\
$B$		& $\chi$			& $\chi^{-1}$	& $1$ \\
$C$		& $\chi$			& $\nu \chi^{-1}$	& $\nu$ \\
$D$		& $1$			& $\nu$	& $\nu$ \\
\bottomrule
\end{tabularx}
\end{table}

%%%%

\subsection{Quasi-static Stirling cycle}
\label{sec:quasiS}

For later reference, we first study the quasi-static limit of the designed cycle. Therein, the system is always at equilibrium and the time required to sweep the cycle is infinite. The analysis of isothermal and isochoric quasi-static processes discussed in the previous section allows us to directly evaluate work, heat and energy increment over each branch. The obtained values are collected in Table \ref{table:QS_energetics}. The total quasi-static work corresponding is 
\begin{equation}
	\mathcal{W}^{QS}\equiv \frac{1-\nu}{2}\ln\chi<0.\label{qs_work_cyle}
\end{equation}
\begin{table}[H] 
\caption{Quasi-static energetics of the Stirling cycle.
	\label{table:QS_energetics}}
\newcolumntype{C}{>{\centering\arraybackslash}X}
\begin{tabularx}{\textwidth}{CCCC}
\toprule
\,	& $\mathcal{W}_{i\to f}^{QS}$	& $\mathcal{Q}_{i\to f}^{QS}$ & $\Delta\mathcal{E}_{i\to f}^{QS}$ \\
\midrule
\end{tabularx}
\begin{tabularx}{\textwidth}{C|CCC}
$(1)\quad A\to B$&$\dfrac{1}{2}\ln\chi$&$-\dfrac{1}{2}\ln\chi$&0\\
$(2)\quad B\to C$&0&$\vphantom{\dfrac{1}{2}} \nu -1$&$\nu-1$\\
$(3)\quad C\to D$&$-\dfrac{\nu}{2}\ln\chi$&$\dfrac{\nu}{2}\ln\chi$&0\\
$(4)\quad D\to A$&0&$\vphantom{\dfrac{1}{2}}1-\nu$&$1-\nu$\\
\midrule
Total		&$\dfrac{1-\nu}{2}\ln\chi$&$\dfrac{\nu-1}{2}\ln\chi$&0 \\
\bottomrule
\end{tabularx}
\end{table}

Since the required time for this engine to operate is infinite, it does not deliver any power.  Nonetheless, the efficiency of such device attains the Carnot value, \begin{equation}
	\eta^{QS} = \eta_C\equiv 1-\frac{\theta_c}{\theta_h}=1-\nu,\label{effQS}
\end{equation}
which is the maximum achievable thermal efficiency, as stated by Carnot's theorem, which is derived as a consequence of the second law of thermodynamics \cite{callen}.

The projection of the quasi-static Stirling cycle onto the $(\kappa,y)$ plane in phase space is illustrated on the left panel of Fig.~\ref{fig:proyeccioncicloStirling}, for the particular choice of parameters $\nu=\chi=0.5$.

%%%

\section{Optimal irreversible Stirling cycle}
\label{sec:opt}

Let us consider now the irreversible version of the above described stochastic Stirling cycle. In contrast to the Carnot-like heat engine, where the irreversibility adds degrees of freedom to the operating points \cite{plata_building_2020}, here the irreversible Stirling cycle is still fully defined by the same parameters  $(\nu,\chi)$  of the quasi-static case.  The four irreversible branches are different from those of the quasi-static case---as depicted in the right panel of Fig.~\ref{fig:proyeccioncicloStirling}. The irreversible branches are swept in a finite time and the heat engine thus delivers a non-zero output power. 

As anticipated in the previous sections, we aim at building the optimal irreversible cycle, in the sense of maximising the delivered power. First, we derive the optimal cycle for given $(\nu,\chi)$, i.e. fixed operating points, and fixed temperature bounds $(\theta_{\min},\theta_{\max})$ in Sec.~\ref{sec:opt-cycle-fixed-points}. Second, we further optimise the cycle with respect to the operating points and the temperature bounds in Sec.~\ref{sec:further-opt}.

\subsection{Optimal cycle for fixed operating points and temperature bounds}\label{sec:opt-cycle-fixed-points}

Henceforth, our objective is to maximise the output power of the cycle, which is given by Eq.~\eqref{power}. Inasmuch as the isochoric branches only contribute to the delivered power through their time spans, we must minimise their duration in order to achieve our goal. We have already addressed this problem in detail in Sec.~\ref{subsect:optimal_isoch} of the previous section. Therein, we obtained that the optimal temperature protocol is of bang-bang type, and it consists of applying the minimal bath temperature in the cooling isochore (i.e. $B\to C $) and the maximum bath temperature in the heating isochore (i.e. $D\to A$).
\begin{figure}
	\begin{subfigure}{.5\textwidth}
		\centering
		\includegraphics[width=0.8\linewidth]{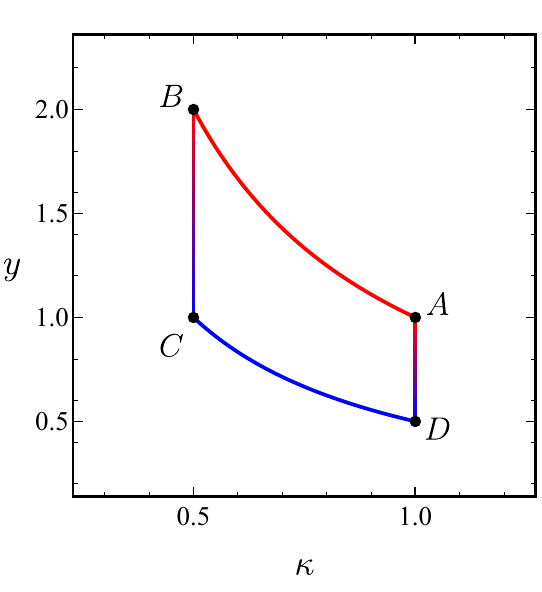}
		\label{fig:sfig1}
	\end{subfigure}%
	\begin{subfigure}{.5\textwidth}
			\raggedright
		\includegraphics[width=0.8\linewidth]{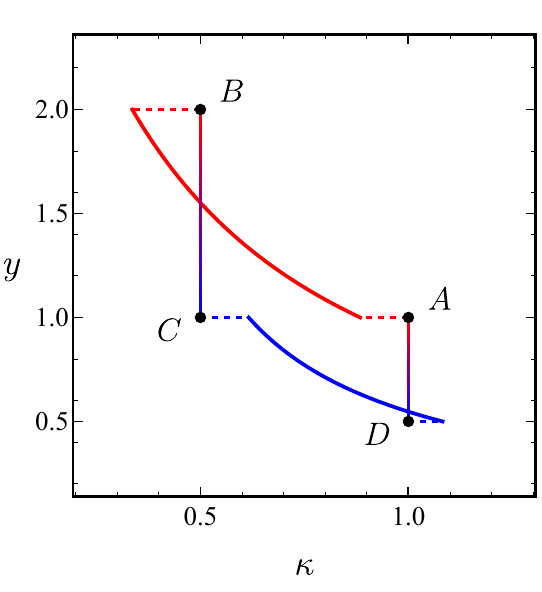}
		\label{fig:sfig2}
	\end{subfigure}
	\caption{Projection onto the $(\kappa, y)$ plane of the phase trajectory in the Stirling heat engine. The left (right) panel corresponds to the reversible (irreversible) cycle. In both panels, $\nu=\chi=0.5$; for the irreversible case, we have considered ideal bounds for the temperature, i.e. $\theta_{\min}\to 0^+$, $\theta_{\max}\to\infty$.}
	\label{fig:proyeccioncicloStirling}
\end{figure}

The above line of reasoning leads to the following optimal times for the isochores:
\begin{align}	&\widetilde{\tau}_{BC}(\theta_{\min})=\frac{1}{2\chi}                     \ln\left(\frac{1-\theta_{\min}}{\nu-\theta_{\min}}\right),           
&\widetilde{\tau}_{DA}(\theta_{\max})=\frac{1}{2}\ln\left(\frac{\theta_{\max}-\nu}{\theta_{\max}-1}\right),	\label{optimal_times_cycle}
\end{align}
which are directly obtained from Eq.~\eqref{optimal_time_brachistochrone}, for the branches $B\to C $ and $D\to A $, respectively. Here, we have explicitly showed the dependence of these optimal times on the temperature bounds.  Note that
  \begin{equation}\label{eq:temp-ordering}
	\theta_{\min}<\theta_c=\nu<\theta_h=1<\theta_{\max}
\end{equation}
in our dimensionless units: the temperatures over the isotherms must be within the interval of admissible values, which ensures that both $\widetilde{\tau}_{BC}$ and $\widetilde{\tau}_{DA}$ are non-negative. The total optimal time for the isochores is
\begin{equation}
	\widetilde{\tau}_{\text{isoc}}(\theta_{\min},\theta_{\max})\equiv\widetilde{\tau}_{BC}(\theta_{\min})+\widetilde{\tau}_{DA}(\theta_{\max}).
\end{equation} 

Regarding the isothermal branches, we need to maximise the work performed by the system $-\mathcal{W}$, i.e. minimise $\mathcal{W}$. In Sec.~\ref{subsect:optimal_isothermal}, we revisited the corresponding optimal protocol for arbitrary time duration of the isotherms \cite{schmiedl_efficiency_2008,plata_optimal_2019}, in which the stiffness $\kappa$ has finite jump discontinuities at both the initial and final times. This behaviour is illustrated on the right panel in Fig.~\ref{fig:proyeccioncicloStirling}, where the dashed lines represent the aforementioned jumps. The optimal work over the isotherms as a function of their time span is directly obtained from Eq.~\eqref{optimal_isothermal_work}: 
\begin{align}
  &\widetilde{\mathcal{W}}_{AB}(\tau_{AB})=
    \frac{1}{2}\ln\chi+\frac{1}{\tau_{AB}}
    \left(\frac{1}{\sqrt{\chi}}-1\right)^2,\\	&\widetilde{\mathcal{W}}_{CD}(\tau_{CD})=                                                                -\frac{\nu}{2}\ln\chi+\frac{\nu}{\tau_{CD}}\left(1-\frac{1}{\sqrt{\chi}}\right)^2.
\end{align}

Since we are considering both $(\kappa,\chi)$ and $(\theta_{\min},\theta_{\max})$ fixed, the minimum times for the isochores are also fixed, whereas the optimal work for the isotherms only depend on their respective times. Now, we may ask what times $\widetilde{\tau}_{AB},\,\widetilde{\tau}_{CD}$ make the power maximum:
\begin{equation}
	\widetilde{\mathcal{P}}=\max_{\tau_{AB},\tau_{CD}}\mathcal{P}\left(\tau_{AB},\tau_{CD}\right),\label{power_optimisation_prolem}
\end{equation}
where 
\begin{equation}
	\mathcal{P}\left(\tau_{AB},\tau_{CD}\right)\equiv- \frac{\widetilde{\mathcal{W}}_{AB}(\tau_{AB})+\widetilde{\mathcal{W}}_{CD}(\tau_{CD})}{\tau_{AB}+\widetilde{\tau}_{BC}+\tau_{CD}+\widetilde{\tau}_{DA}}=- \frac{\widetilde{\mathcal{W}}_{AB}(\tau_{AB})+\widetilde{\mathcal{W}}_{CD}(\tau_{CD})}{\tau_{AB}+\widetilde{\tau}_{\text{isoc}}+\tau_{CD}}.\\\label{p_as_fuction_of_isothermal_times}
\end{equation}
To answer this question, it is useful to rewrite Eq.~\eqref{p_as_fuction_of_isothermal_times} as follows,
\begin{equation}
	\mathcal{P}\left(\tau_{AB},\tau_{CD}\right)=-\frac{1}{\tau_{\text{cyc}}}\left[\mathcal{W}^{QS}+\alpha\left(\frac{1}{\tau_{AB}}+\frac{\nu}{\tau_{CD}}\right)\right],
\end{equation}
where $\mathcal{W}^{QS}$ is the total work corresponding to the quasi-static cycle, given by Eq.~\eqref{qs_work_cyle},
\begin{equation}
	\tau_{\text{cyc}}\equiv\tau_{AB}+\widetilde{\tau}_{\text{isoc}}+\tau_{CD}
\end{equation}
is the total duration of the cycle, and the coefficient $\alpha$ is defined as
\begin{equation}
    \alpha\equiv \left(\frac{1}{\sqrt{\chi}}-1\right)^2,
\end{equation}
for the sake of compactness.

We maximise the power by imposing that the partial derivatives of $\mathcal P$ with respect to $\tau_{AB}$ and $\tau_{CD}$ be equal to zero. A quadratic equation is found, which only has one physical solution, with $\widetilde\tau_{AB}>0$ and $\widetilde\tau_{CD}>0$:
\begin{equation}
  \widetilde\tau_{AB}=-\frac{\alpha}{\mathcal{W}^{QS}}\left(1+\sqrt{\nu}\right)\left(1+\sigma\right),
  \quad \widetilde{\tau}_{CD}=\widetilde{\tau}_{AB}\sqrt{\nu},
\label{opt_times_AB_CD}
\end{equation}
where we have employed the definition of a new parameter
\begin{equation}
  \sigma \equiv\sqrt{1- \frac{\mathcal{W}^{QS}}{\alpha\left(1+\sqrt{\nu}\right)^{2}} \widetilde\tau_{\text{isoc}}}>1.
  \label{w_and_sigma_defs}
\end{equation}

Therefore, we have solved the optimisation problem presented in Eq.~\eqref{power_optimisation_prolem} and the optimal irreversible Stirling cycle is fully characterised for any given operation points and extremal bath temperatures $(\theta_{\min},\,\theta_{\max})$.  The corresponding protocols for the trap stiffness $\kappa(\tau)$ in the isothermal branches and the bath temperature $\theta(\tau)$ in the isochoric connections have been described in detail in section \ref{sec:building_blocks_for_a_Stirling_cycle}.

The energetics of the designed optimal irreversible Stirling cycle can be readily calculated. In analogy with Table \ref{table:QS_energetics}, which collected the energetic description of the quasi-static limit, we present the corresponding description for the optimal irreversible cycle in Table \ref{table:irr_energetics}. Note that the total work in the irreversible case is
\begin{equation}
  \widetilde{\mathcal{W}}=
  \dfrac{\sigma}{1+\sigma}\mathcal{W}^{QS}.\label{opt_work}
\end{equation}
This result exhibits a strong parallelism to that in Ref.~\cite{plata_building_2020} for an optimal irreversible Carnot engine. As pointed out therein, the form of Eq.~\eqref{opt_work} yields a physical interpretation for the parameter $\sigma$: it measures the deviation of the total irreversible work from the value corresponding to the quasi-static case. In the limit $\sigma\to\infty$, one has $	\widetilde{\mathcal{W}}\to\mathcal{W}^{QS}$. From the definition of $\sigma$ in Eq.~\eqref{w_and_sigma_defs}, it is clear that the limit $\sigma\to\infty$ corresponds, indeed, to infinitely slow isochoric processes: $\widetilde\tau_{\text{isoc}}\to\infty$. Consistently, Eq.~\eqref{opt_times_AB_CD} evinces that the duration of the optimal isothermal branches also diverge in the limit $\sigma\to\infty$.
\begin{table}[H] 
\caption{Energetics of the optimal irreversible Stirling cycle.
	\label{table:irr_energetics}}
\newcolumntype{C}{>{\centering\arraybackslash}X}
\begin{tabularx}{\textwidth}{CCCC}
\toprule
\,	& $\widetilde{\mathcal{W}}_{i\to f}$	& $\widetilde{\mathcal{Q}}_{i\to f}$ & $\Delta\widetilde{\mathcal{E}}_{i\to f}$ \\
\midrule
\end{tabularx}
\begin{tabularx}{\textwidth}{C|CCC}
$(1)\quad A\to B$&$\dfrac{\sqrt{\nu}+\sigma}{\left(1-\nu\right)\left(1+\sigma\right)}\mathcal{W}^{QS}$&$-\dfrac{\sqrt{\nu}+\sigma}{\left(1-\nu\right)\left(1+\sigma\right)}\mathcal{W}^{QS}$&0 \\
$(2)\quad B\to C$&0&$\nu -1$&$\nu-1$\\
$(3)\quad C\to D$&$-\dfrac{\sqrt{\nu}\left(1+\sqrt{\nu}\sigma\right)}{\left(1-\nu\right)\left(1+\sigma\right)}\mathcal{W}^{QS}$&$\dfrac{\sqrt{\nu}\left(1+\sqrt{\nu}\sigma\right)}{\left(1-\nu\right)\left(1+\sigma\right)}\mathcal{W}^{QS}$&0\\
$(4)\quad D\to A$&0&$1-\nu$&$1-\nu$\\
\midrule
Total		&$\dfrac{\sigma}{1+\sigma}\mathcal{W}^{QS}$&$-\dfrac{\sigma}{1+\sigma}\mathcal{W}^{QS}$&0\\
\bottomrule
\end{tabularx}
\end{table}

Therefore, the optimal power for given operating points, as defined by $(\nu,\chi)$, and limiting temperatures $(\theta_{\min},\theta_{\max})$ is
\begin{equation}
  \widetilde{\mathcal{P}}\left(\nu,\chi;\,\theta_{\min},\theta_{\max}\right)= - \frac{\left[ \mathcal{W}^{QS}(\nu,\chi)\right]^2}{\alpha(\chi)\left(1+\sqrt{\nu}\right)^{2}\left[1+\sigma(\nu,\chi ;\theta_{\min}, \theta_{\max})\right]^{2}},
  \label{power4parameters}
\end{equation}
where we explicitly show the dependence on the system parameters of the different quantities involved. The associated efficiency at maximum power is obtained by particularising Eq.~\eqref{eq:eta-func-W-Q} to the optimal cycle:
\begin{equation}
  \widetilde{\eta}(\nu,\chi;\theta_{\min},\theta_{\max})=
  1-\frac{\widetilde{\mathcal{W}}_{CD}}{\widetilde{\mathcal{Q}}_{AB}}=
  1+\frac{\widetilde{\mathcal{W}}_{CD}}{\widetilde{\mathcal{W}}_{AB}}
  =\underbrace{1-\nu}_{\eta_{C}}-\underbrace{\frac{\sqrt{\nu}\left(1-\nu\right)}{\sqrt{\nu}+\sigma(\nu,\chi ;\theta_{\min}, \theta_{\max})}}_{>0}.
  \label{opt_irrev_efficiency}
\end{equation}
Therefore, the Carnot efficiency $\eta_C=1-\nu$ is an upper bound for $\widetilde{\eta}$, and this bound is only attained in the quasi-static limit $\sigma\to\infty$, as expected. 

Equation~\eqref{opt_irrev_efficiency} can also be used to show that the Curzon-Ahlborn efficiency \cite{curzon_efficiency_1975}
\begin{equation}
  \eta_{CA}\equiv 1-\sqrt{\nu}
\end{equation}
is not an upper but a lower bound for the efficiency of the optimal irreversible Stirling cycle. Indeed, after some calculations, it is possible to write
\begin{equation}
  \widetilde{\eta}
  =\eta_{CA}+ \underbrace{\frac{\sqrt{\nu}\left(1-\sqrt{\nu}\right)
      \left[\sigma(\nu,\chi ;\theta_{\min}, \theta_{\max})-1 \right]}{\sqrt{\nu}+\sigma(\nu,\chi ;\theta_{\min}, \theta_{\max})}}_{>0}.
  \label{eff_greater_than_CA}
\end{equation}
Our optimal Stirling-like engine is thus found to operate always above the Curzon-Ahlborn efficiency at its maximum power.

Curzon and Ahlborn found $\eta_{CA}$ to be the efficiency of a Carnot engine operating at maximum power when limitations in the rates of heat transfer were considered \cite{curzon_efficiency_1975}. This result was derived for an endoreversible heat engine, generating a long-standing debate about its universality as an upper bound for the efficiency at maximum power. Its generality as an upper bound has been discarded, since efficiencies at maximum power below and above $\eta_{CA}$ have been reported in the literature \cite{esposito_universality_2009, schmiedl_efficiency_2008, esposito_efficiency_2010, plata_building_2020}.

\begin{figure}
	\begin{subfigure}{.5\textwidth}
		\raggedright
		\includegraphics[width=0.99\linewidth]{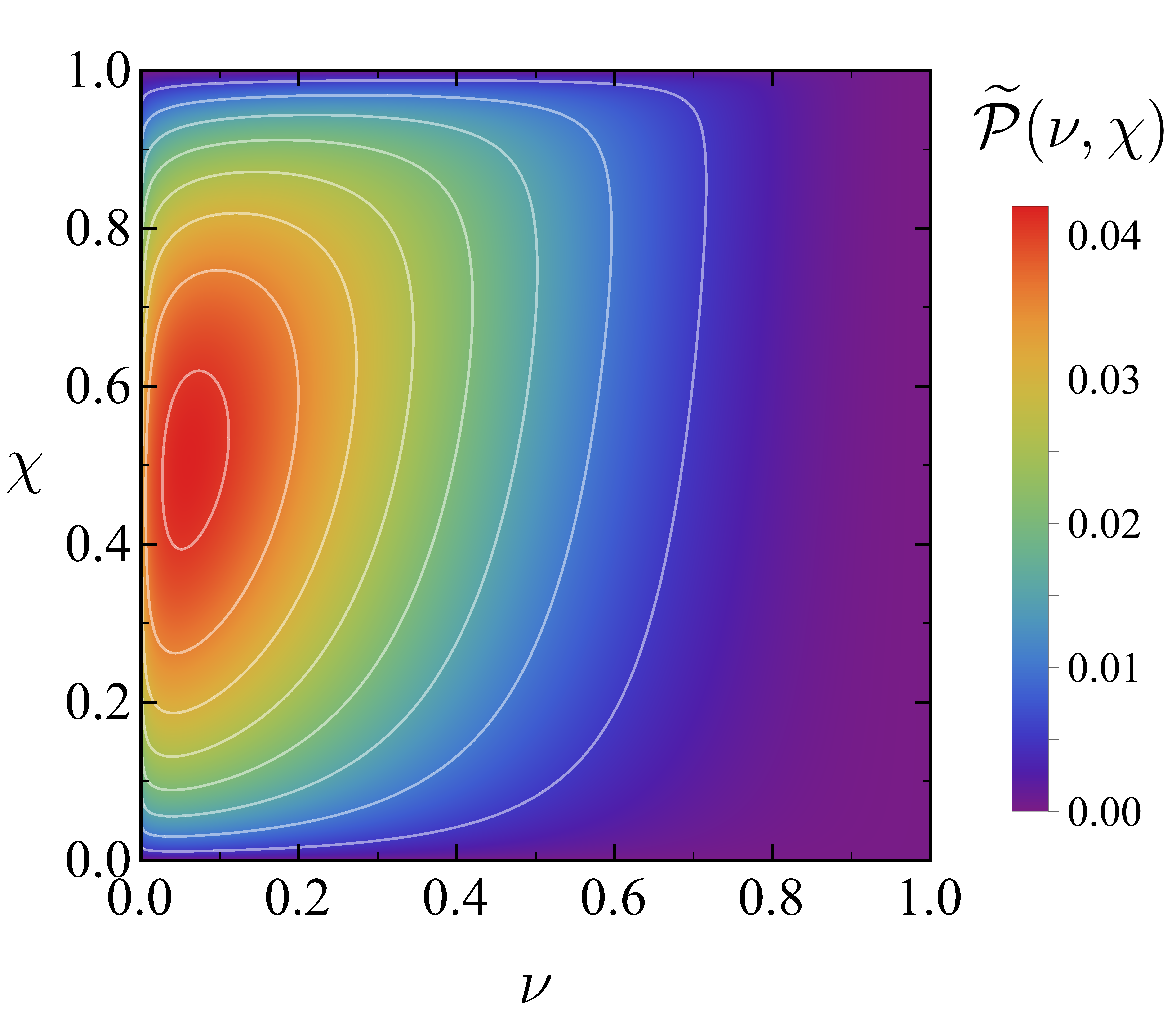}
	\end{subfigure}%
	\begin{subfigure}{.5\textwidth}
		\raggedleft
		\includegraphics[width=0.99\linewidth]{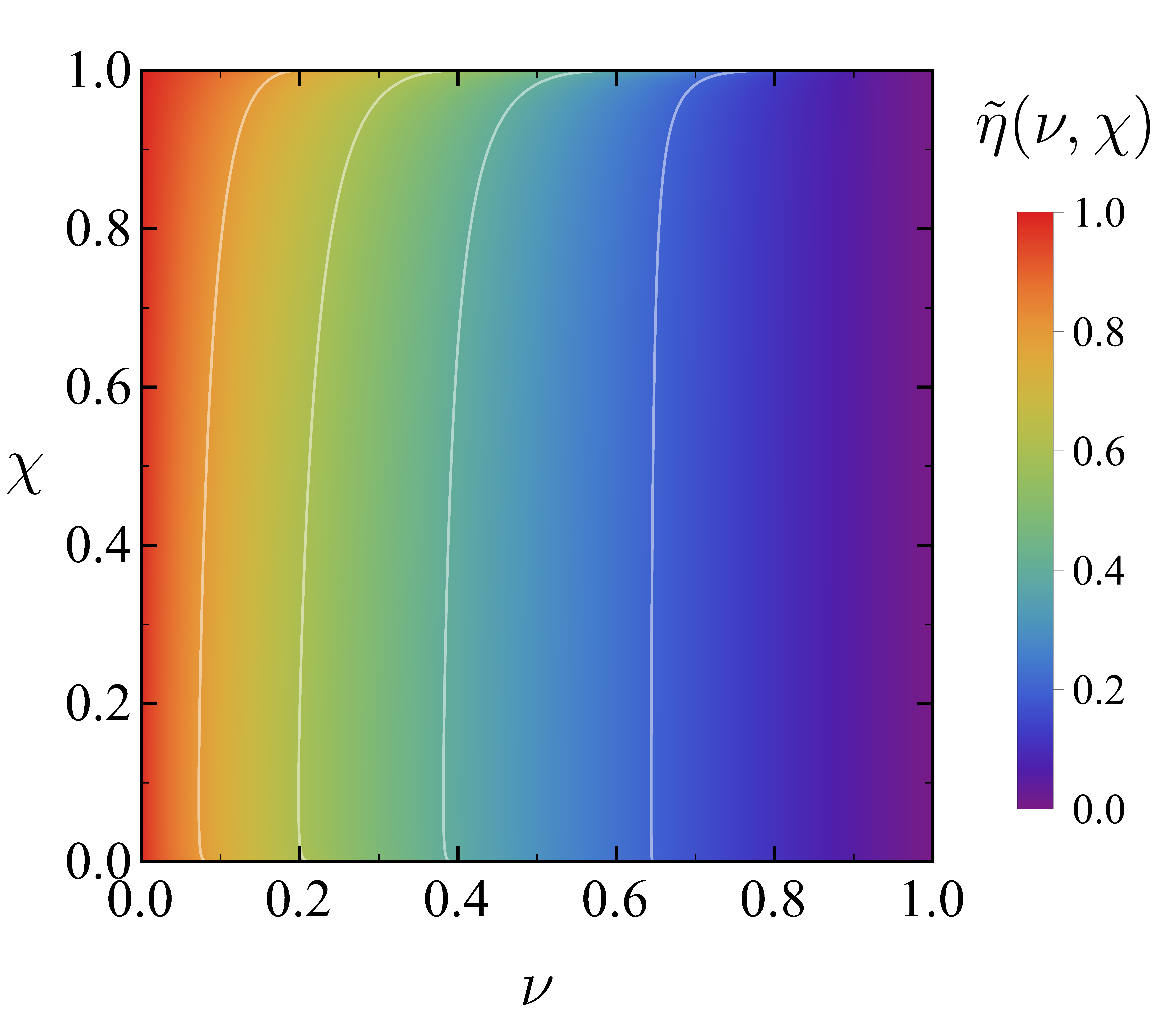}
	\end{subfigure}
	\caption{Density plots of the optimal power (left) and the corresponding efficiency (right) in the $(\nu,\chi)$ plane. We have considered the loosest bounds for the temperature: $\theta_{\min}\to0^+$ and $\theta_{\max}\to\infty$.}
	\label{fig:power_and_efficiency_DPs}
\end{figure}

 To illustrate the results obtained in this section, density plots of the maximum power $\widetilde{P}$ and the corresponding efficiency $\widetilde{\eta}$ are presented in Fig.~\ref{fig:power_and_efficiency_DPs}  as a function of $(\nu,\chi)$, in the ideal case of $\theta_{\min}\to0^+$, $\theta_{\max}\to\infty$. It is worth stressing that the order of magnitude of $\widetilde{P}$ is  $10^{-2}$.  The experimental realization of a a Stirling engine in Ref.~\cite{blickle_realization_2012} reported a delivered power of the order of $10^{-4}$---after translating the results therein to our dimensionless variables. Therefore, our optimal Stirling heat engine gives a $10^{2}$ improvement factor, in terms of extracted output power.\footnote{For the estimation of the delivered power in the experiment reported in Ref.~\cite{blickle_realization_2012}, we have taken into account that the typical drag coefficient of a micrometre-sized bead in water solution is around $\lambda= 10^{-8}\; \text{kg}\; \text{s}^{-1}$, whereas the stiffness in that experiment was around $k=10^{-6} \;\text{kg} \; \text{s}^{-2}$ and the cold and hot bath temperature were $22\;^\circ$C and $86\;^\circ$C.}

\subsection{Further optimisation of the irreversible Stirling cycle}\label{sec:further-opt}

Hitherto, we have studied the optimisation of the irreversible Stirling cycle for fixed operating points, i.e. given values of $(\nu,\chi)$, and fixed limiting temperatures $(\theta_{\min},\theta_{\max})$. The question of what choice of these parameters renders maximal power naturally arises, and this section is devoted to answering this question. In particular, we investigate the optimal values of the compression and temperature ratios $(\chi,\nu)$. First, we consider the case of ideal temperature bounds $(\theta_{\min}\to 0^{+},\theta_{\max}\to +\infty)$ in Sec.~\ref{sec:opt-ideal-case}. Second, we investigate more realistic bounds for the temperature in Sec.~\ref{sec:realistic-temp-bounds}.

%%%%%

\subsection{Further optimisation with ideal temperature bounds}\label{sec:opt-ideal-case}

Here, we consider ideal temperature bounds, i.e. the temperature has no upper bound, $\theta_{\text{max}}\to \infty$, and the lower bound corresponds to absolute zero, $\theta_{\text{min}}\to 0^+$. The optimal power for given operating points thus depends only on $(\nu,\chi)$, $\widetilde{\mathcal{P}}=\widetilde{\mathcal{P}}(\nu,\chi)$, and has already been presented in the left panel of Fig.~\ref{fig:power_and_efficiency_DPs}. Therein, it is clearly observed that $\widetilde{\mathcal{P}}$ for fixed $\nu$ vanishes both for $\chi\to 0^{+}$ and $\chi\to 1^{-}$, which implies that there appears a maximum of $\widetilde{\mathcal{P}}$ as a function of $\chi$, at fixed $\nu$.\footnote{In the limit $\chi\to 0^{+}$, the minimum time over the isochores diverge, whereas in the limit $\chi\to 1^{-}$, the maximum work vanishes.} This motivates the two-step procedure followed below to find the overall maximum of $\widetilde{\mathcal{P}}$.

The optimisation of the power over $(\chi,\nu)$ is carried out in two steps. First, we look into its optimisation over the compression ratio $\chi$, for fixed temperature ratio $\nu$, i.e. we look for
\begin{equation}
  \widetilde{\mathcal{P}}^*(\nu)\equiv
  \max_{\chi\in(0,1)}\widetilde{\mathcal{P}}\left(\nu,\chi\right).
\end{equation}
We denote by $\chi^*(\nu)$ the value of the compression ratio at which the maximum of $\widetilde{\mathcal{P}}\left(\nu,\chi\right)$ is attained for a given value of $\nu$. So we can write 
\begin{equation}
\widetilde{\mathcal{P}}^*(\nu)=\widetilde{\mathcal{P}}\left(\nu,\chi^*(\nu)\right).
\end{equation}
Second, we seek the overall optimal Stirling cycle by maximising $\widetilde{\mathcal{P}}^*(\nu)$ over the temperature ratio $\nu$, i.e. we look for
\begin{equation} 	\widetilde{\mathcal{P}}^{**}\equiv\max_{\substack{\chi\in(0,1)\\\nu\in(0,1) }}\widetilde{\mathcal{P}}\left(\nu,\chi\right)= \max_{\nu\in(0,1)}\widetilde{\mathcal{P}}^*(\nu).
\end{equation}
We denote by $\nu^{*}$ the temperature ratio at which $\widetilde{\mathcal{P}}^*(\nu)$ reaches its maximum. Moreover, we define $\chi^{**}\equiv \chi^{*}(\nu^{*})$ to write
\begin{equation} 	\widetilde{\mathcal{P}}^{**}=\widetilde{\mathcal{P}}^*(\nu^{*})=\widetilde{\mathcal{P}}\left(\nu^*,\chi^{**}\right). 
\end{equation}
This maximisation over $(\chi,\nu)$ has to be performed numerically,  since the involved formula for $\widetilde{\mathcal{P}}\left(\nu,\chi \right)$ does not allow us to obtain closed-form expressions for the optimal values.

\begin{figure}%[h!]
	\centering
	\captionsetup{width=.85\linewidth}
	\includegraphics[width=1 \textwidth]{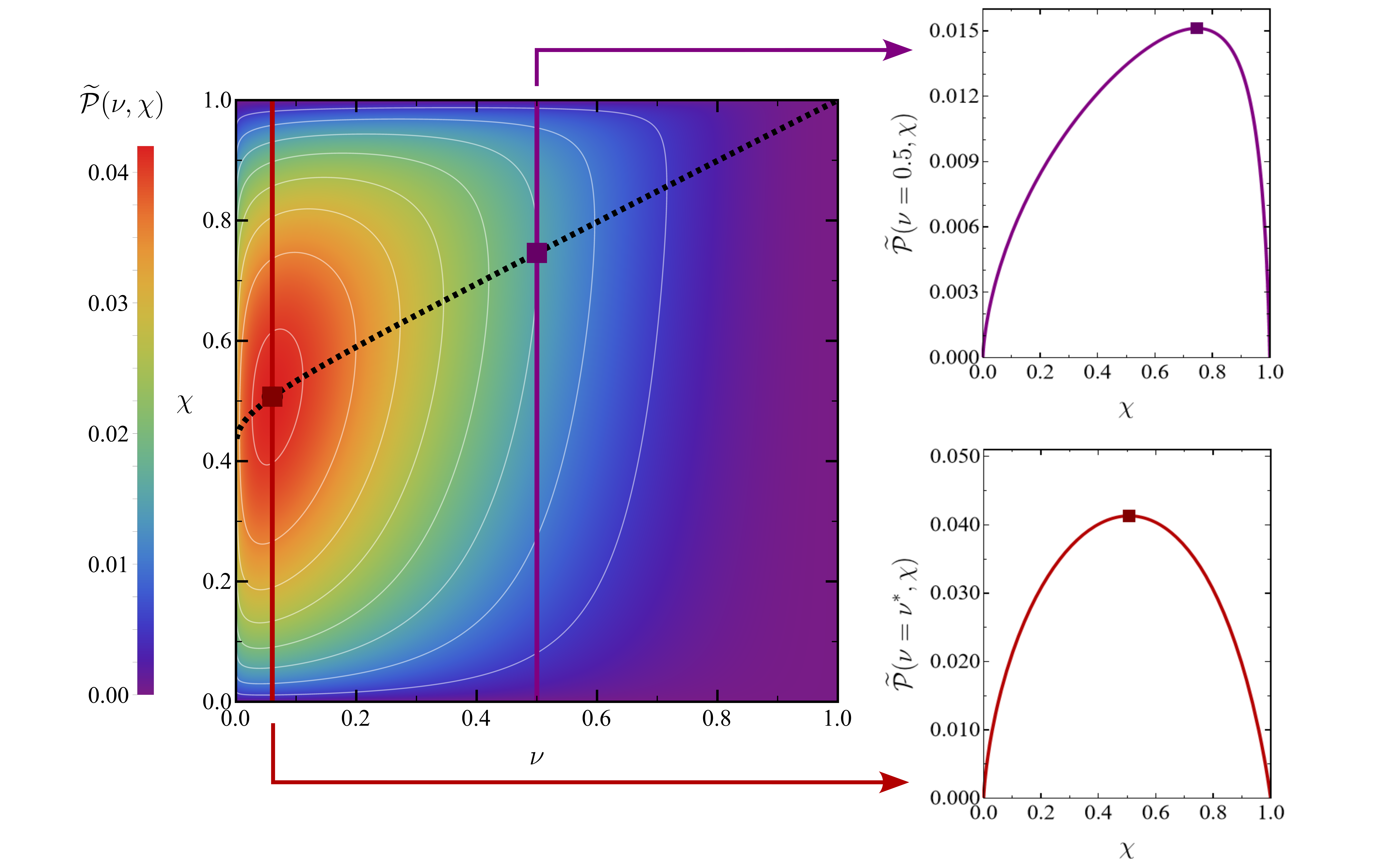}
	\caption{Density plot of the optimal power in the $(\nu,\chi)$ plane (left) and vertical sections for fixed values of the temperature ratio $\nu$ (right). The curve $\chi = \chi^*(\nu)$ (dotted line) gives the compression ratio yielding optimal power for every temperature ratio. On the right, the upper panel corresponds to $\nu=0.5$ and the bottom one to $\nu=\nu^*$. The points at which maximum power is reached in each case are also displayed (squares).}
	\label{fig:optimal_chi_for_fixed_nu}
\end{figure}

To illustrate the numerical optimisation of the power, we present the density plot of the optimal power in the $(\nu,\chi)$ plane in Fig.~\ref{fig:optimal_chi_for_fixed_nu}, together with the curve $\chi = \chi^*(\nu)$---as well as two plots showing the behaviour of the optimal power as a function of the compression ratio $\chi$ for twovalues of the temperature ratio $\nu$. We find $\chi^*(\nu)$ to be monotonically increasing with $\nu$. The overall maximum power is $\widetilde{\mathcal{P}}^{**}\simeq 0.041$, which is attained at $\nu^*\simeq 0.060$ and $\chi^{**}\equiv\chi^*(\nu^*)\simeq 0.507$.

Similarly to the optimal power, the efficiency at maximum power for fixed operating points only depends on $(\nu,\chi)$, $\widetilde{\eta}=\widetilde{\eta}(\nu,\chi)$, and has already been presented in the right panel of Fig.~\ref{fig:power_and_efficiency_DPs}. The efficiencies corresponding to the power optimisation over the compression ratio and to the overall maximum power are denoted in an analogous manner,
\begin{equation}
	\widetilde{\eta}^*(\nu)\equiv\widetilde{\eta}\left(\nu,\chi^*(\nu)\right),
	\quad
	\widetilde{\eta}^{**}\equiv\widetilde{\eta}\left(\nu^*,\chi^{**}\right)\simeq 0.842.
\end{equation}
\begin{figure}
	\centering
	\captionsetup{width=.8\linewidth}
	\includegraphics[width=0.8 \textwidth]{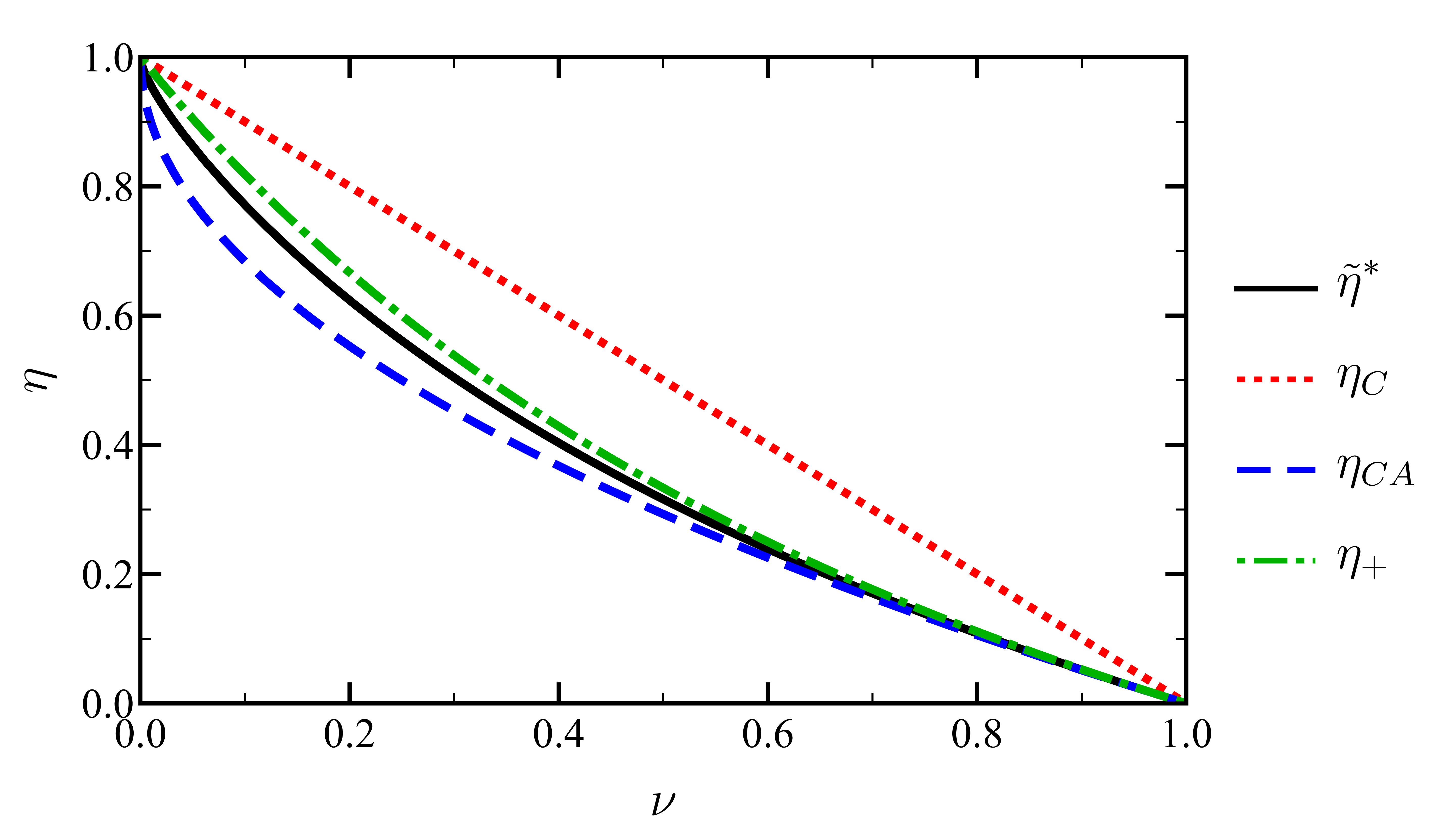}
	\caption{Efficiency at maximum power in our approach (black solid line), Carnot efficiency (red dotted line), Curzon-Ahlborn efficiency (blue dashed line) and $\eta_+$ (green dash-dotted line), defined in Eq.~\eqref{eq:eta+},  as a function of the temperature ratio $\nu$.}
	\label{fig:Eff_IdealTempBounds}
      \end{figure}
      
In the previous section, we proved that the efficiency is always below Carnot's, as expected, but above Curzon-Ahlborn's---for arbitrary values of the temperature and compression ratios $(\nu, \chi)$. Therefore, these bounds also apply after optimising over $\chi$:
\begin{equation}
	\eta_{CA}<\widetilde{\eta}^*<\eta_{C}.
\end{equation}
This behaviour is illustrated in Fig.~\ref{fig:Eff_IdealTempBounds}. Also, it shows that the upper bound at low dissipation of $\eta$ for engines reaching the Carnot efficiency in the reversible limit~\cite{esposito_efficiency_2010}, 
\begin{equation}
\label{eq:eta+}
	\eta_+\equiv\frac{\eta_C}{2-\eta_C}=\frac{1-\nu}{1+\nu}
\end{equation}
holds for our optimal Stirling engine---despite not working in the low dissipation regime.
 
\begin{figure}%[h!]
	\centering
	\captionsetup{width=.8\linewidth}
        \includegraphics[width=0.875 \textwidth]{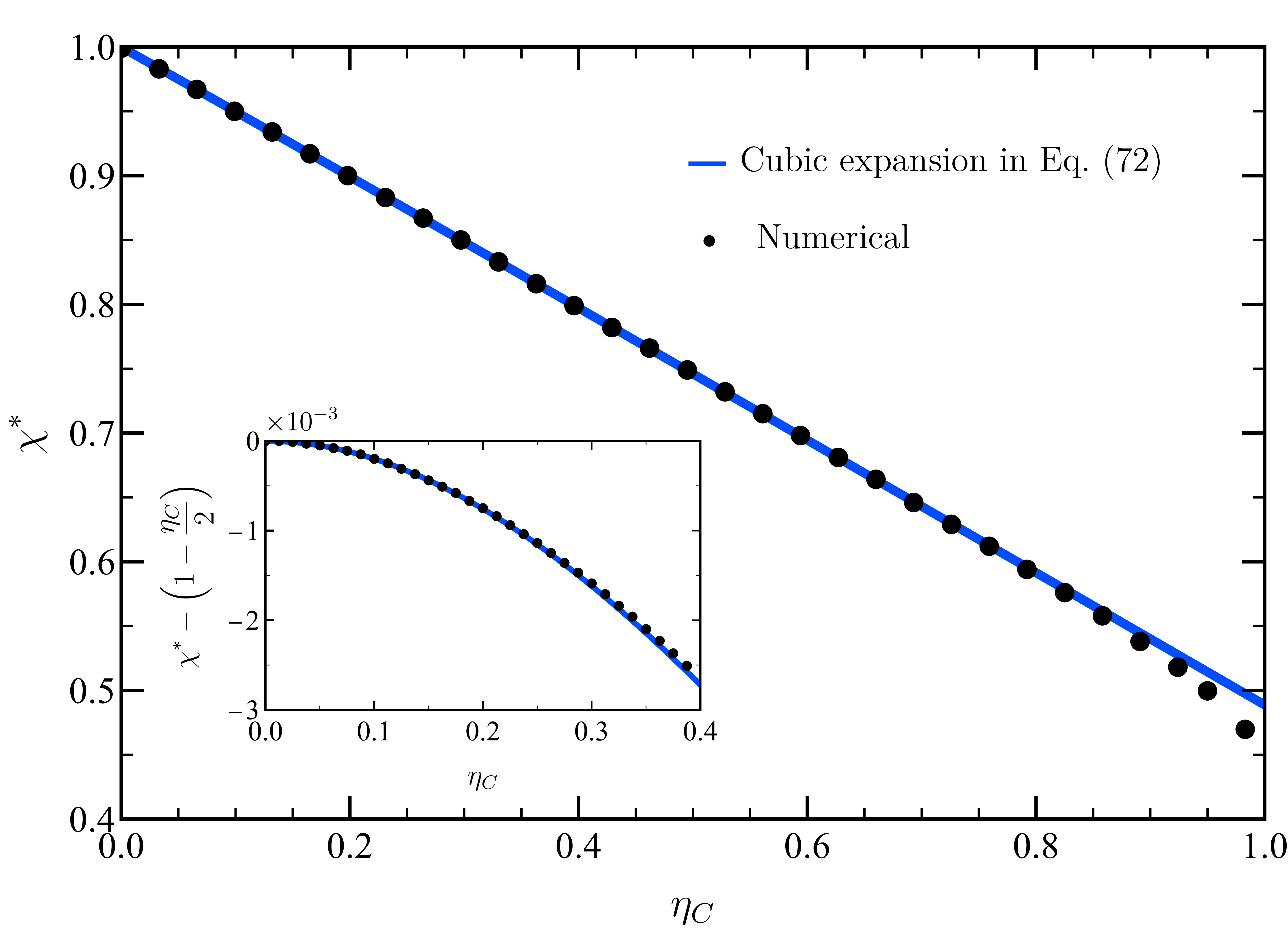}
	\caption{Optimal compression ratio $\chi^*$ as a function of the Carnot efficiency $\eta_C=1-\nu$. We compare numerical results (black circles) with the theoretical expansion in $\eta_C$ up to cubic order, as given by Eqs.~\eqref{eq:chi*-expand} and \eqref{eq:chi*-expand-coef} (blue solid line). The inset shows the difference between the numerical results and the first-order approximation $1+a_{1}\eta_{C}$, which is very small, of the order of $10^{-3}$, up to $\eta_{C}\simeq 0.4$.
        }
	\label{fig:numerical_check}
      \end{figure}

Let us now consider the efficiency at maximum power $\widetilde{\eta}^*$ in the limit $\nu\to 1^{-}$, i.e. $\eta_C=1-\nu \ll 1$. In order to get an analytical expression for $\chi^{*}$ in this regime, our approach is the following: we expand $\chi^*$ in powers of $\eta_C$ and introduce this expansion in $\widetilde{\eta}^*(\nu)=\widetilde{\eta}(\nu,\chi^*(\nu))$ to obtain the power expansion of $\widetilde{\eta}^*$ up to the desired order in $\eta_C$. Specifically, we introduce the following ansatz for $\chi^*$:
 \begin{equation}\label{eq:chi*-expand}
 	\chi^* = 1+a_1\,\eta_C+a_2\,\eta_C^2+a_3\,\eta_C^3+O\left(\eta_C^4\right),
 \end{equation}
where the coefficients $(a_{1},a_{2},a_{3})$ are determined by enforcing the first three terms in the expansion of $\partial\widetilde{\mathcal{P}}/\partial\chi$ to vanish at $\chi^*$, since it corresponds to the optimal output power. The zero-order value of $\chi^*$ stems from having $\chi^*\to1$ for $\nu\to1$, as shown by the dotted line in Fig.~\ref{fig:optimal_chi_for_fixed_nu}. The described procedure yields
\begin{equation}\label{eq:chi*-expand-coef}
a_1 = -\frac12,\quad a_2 = -\frac{1}{48},\quad a_3 = \frac{11}{1152}.
\end{equation}
In Fig.~\ref{fig:numerical_check}, we check the accuracy of this theoretical prediction for $\chi^{*}$, up to third order in $\eta_C$. Its agreement with the numerically found values of $\chi^{*}$ is quite good, even when considering values of $\eta_C$ not so close to zero. The inset shows the deviation from the linear approximation $\chi^{*}=1+a_{1}\eta_{C}$, which is very small---consistently with the smallness of the non-linear coefficients, $a_{2}\simeq 0.021$ and $a_{3}\simeq 0.001$.

The corresponding expansion for the efficiency at maximum power,
\begin{equation}
  \widetilde{\eta}^*=
  \frac{\eta_C}{2}+\frac{3}{16}\eta_C^2+\frac{41}{384}\eta_C^3+
  O\left(\eta_C^4\right).
	\label{exp_eta*}
\end{equation}
follows after inserting the above expansion for $\chi^*$ into $\widetilde{\eta}^*(\nu)\equiv\widetilde{\eta}\left(\nu,\chi^*(\nu)\right)$. Note that the linear  response approximation $\widetilde{\eta}^{*}=\eta_{C}/2$ has been proved to be a general result for the efficiency at maximum power, as a consequence of the Onsager reciprocity theorem---which has been considered as the fourth law of thermodynamics \cite{van_den_broeck_eff}.

The expansion of the Curzon-Ahlborn efficiency in powers of $\eta_{C}$ is 
\begin{equation}\label{eq:eta_CA_expand}
{\eta}_{CA}=\frac{\eta_C}{2}+\frac{\eta_C^2}{8}+\frac{\eta_C^3}{16}+O\left(\eta_C^4\right),
\end{equation}
and the corresponding expansion of the upper bound $\eta_+$ is
\begin{equation}\label{eq:eta+_expand}
{\eta}_{+}=\frac{\eta_C}{2}+\frac{\eta_C^2}{4}+\frac{\eta_C^3}{8}+O\left(\eta_C^4\right).
\end{equation}
As expected, Eqs.~\eqref{exp_eta*}, \eqref{eq:eta_CA_expand}, and \eqref{eq:eta+_expand} coincide to first order in $\eta_C$, due to the universality of the linear response term. Nevertheless, the quadratic terms do not: this is not surprising, since we have proved in the previous section that our efficiency at maximum power lies above $\eta_{CA}$ but below $\eta_{+}$. Consistently, the quadratic coefficient in Eq.~\eqref{exp_eta*} for $\widetilde{\eta}^*$ lies between the corresponding values for $\eta_{CA}$ and $\eta_{+}$: $1/8<3/16<1/4$.

%%%%%

\subsection{Further optimisation for arbitrary temperature bounds $(\theta_{\min}$, $\theta_{\max})$}\label{sec:realistic-temp-bounds}

Now we move on to considering arbitrary bounds $\left(\theta_{\min}, \theta_{\max}\right)$ in the temperature control. Restrictions concerning these parameters and the temperatures of the cold and hot branches of the cycle arise, as expressed by Eq.~\eqref{eq:temp-ordering}.

\begin{figure}
	\begin{subfigure}{.5\textwidth}
		\raggedright
		\includegraphics[width=0.95\linewidth]{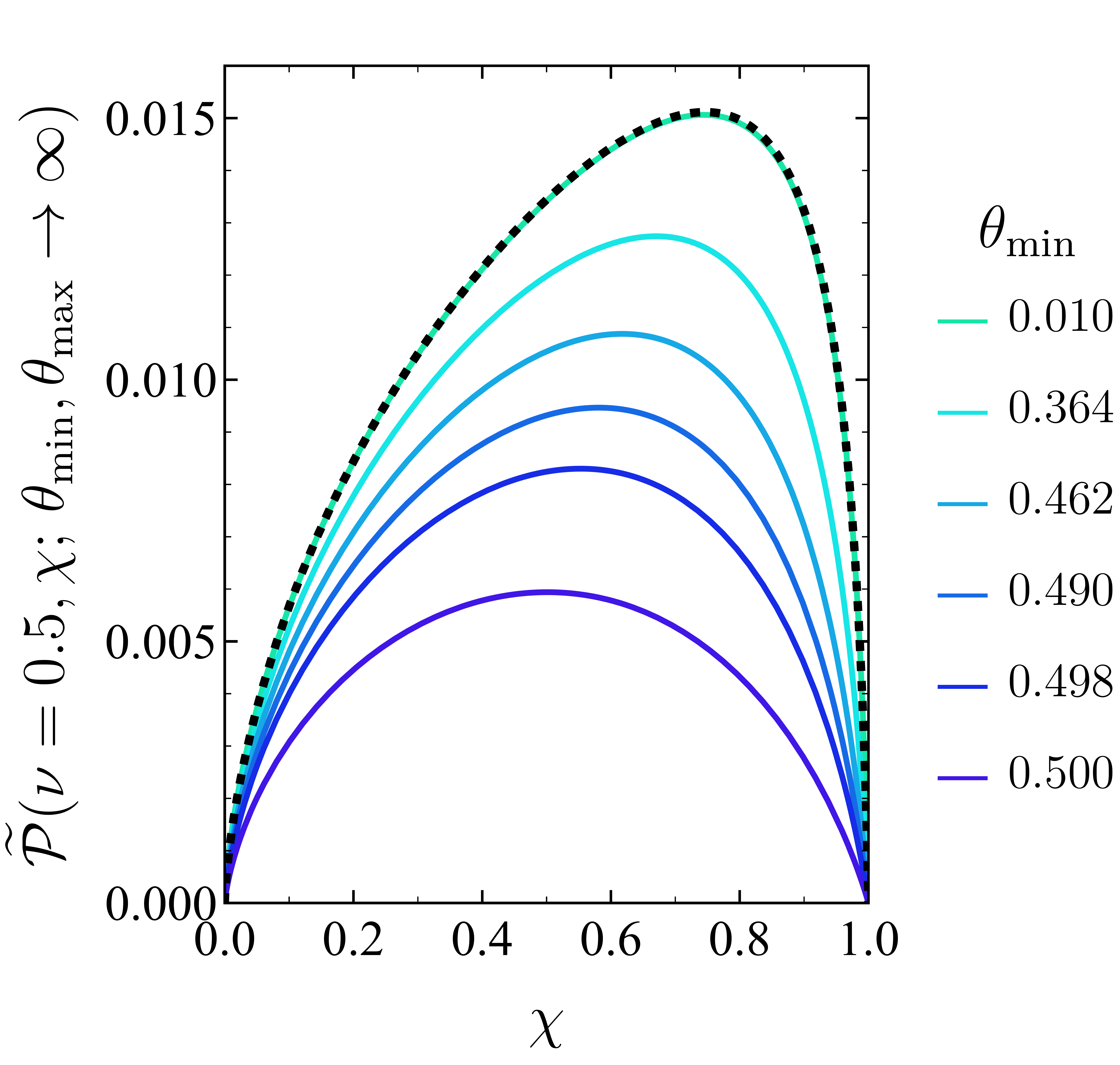}
	\end{subfigure}%
	\begin{subfigure}{.5\textwidth}
		\raggedleft
		\includegraphics[width=0.95\linewidth]{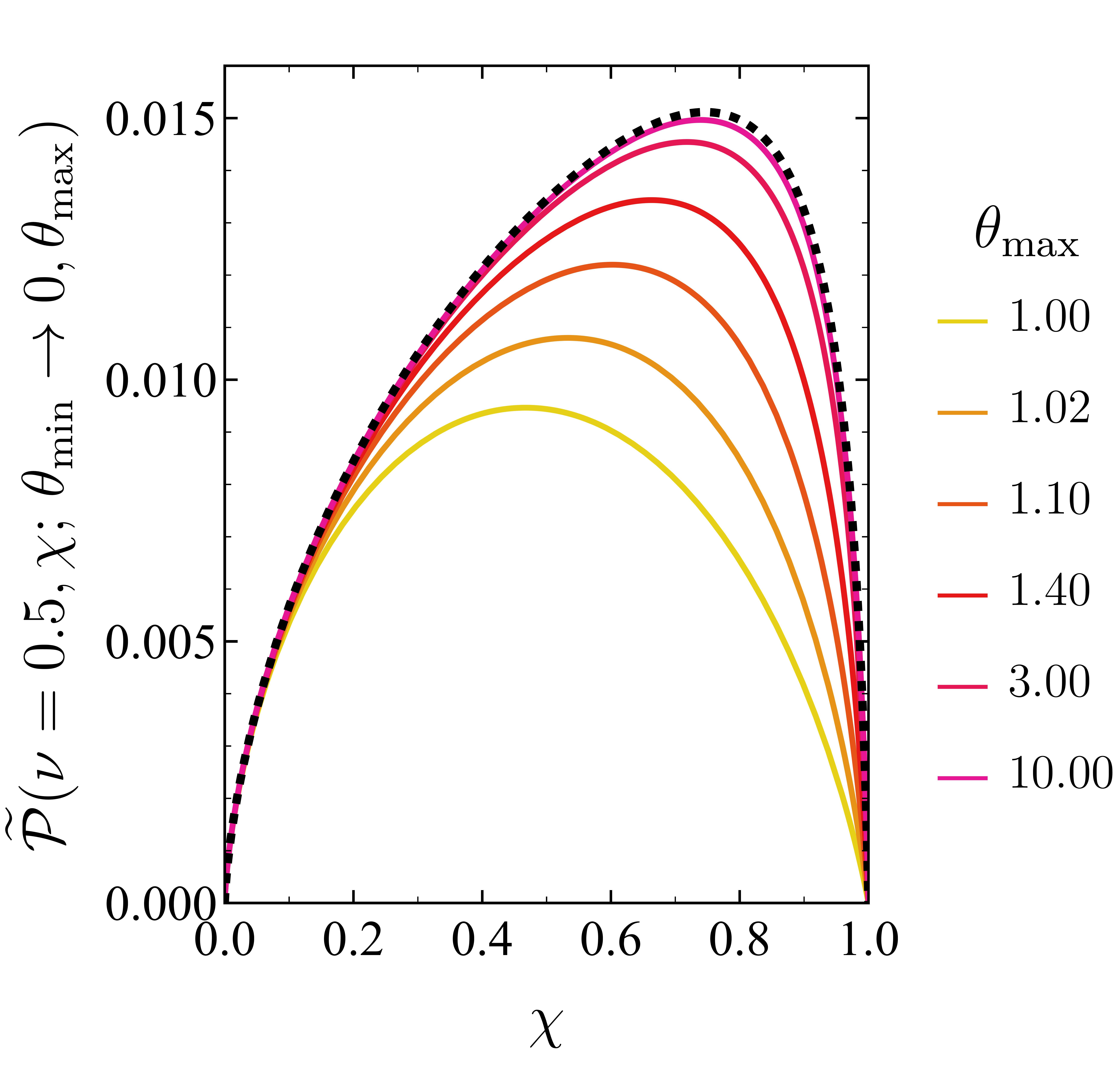}
	\end{subfigure}
	\begin{subfigure}{.5\textwidth}
		\raggedright
		\includegraphics[width=0.95\linewidth]{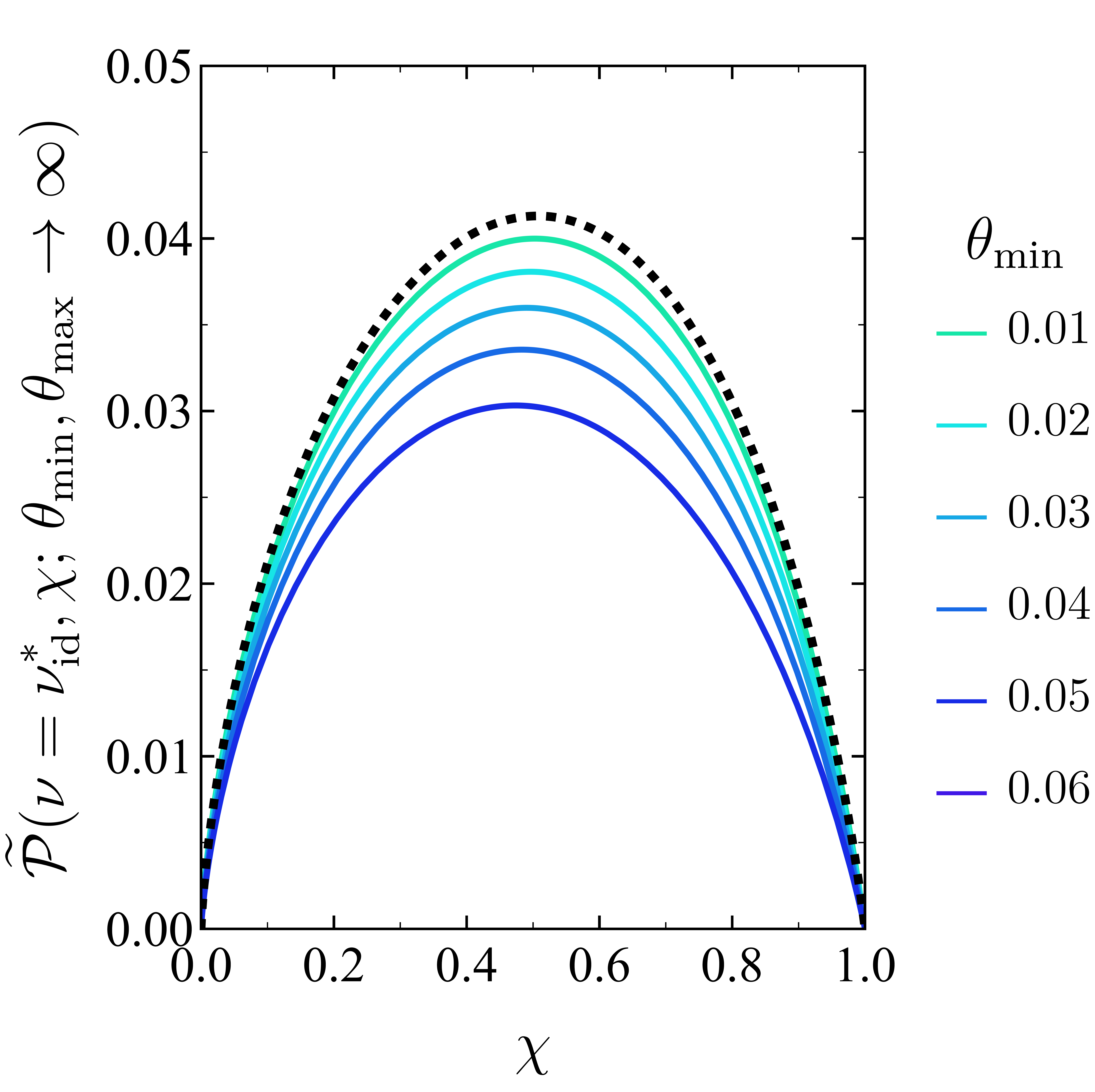}
	\end{subfigure}%
	\begin{subfigure}{.5\textwidth}
		\raggedleft
		\includegraphics[width=0.95\linewidth]{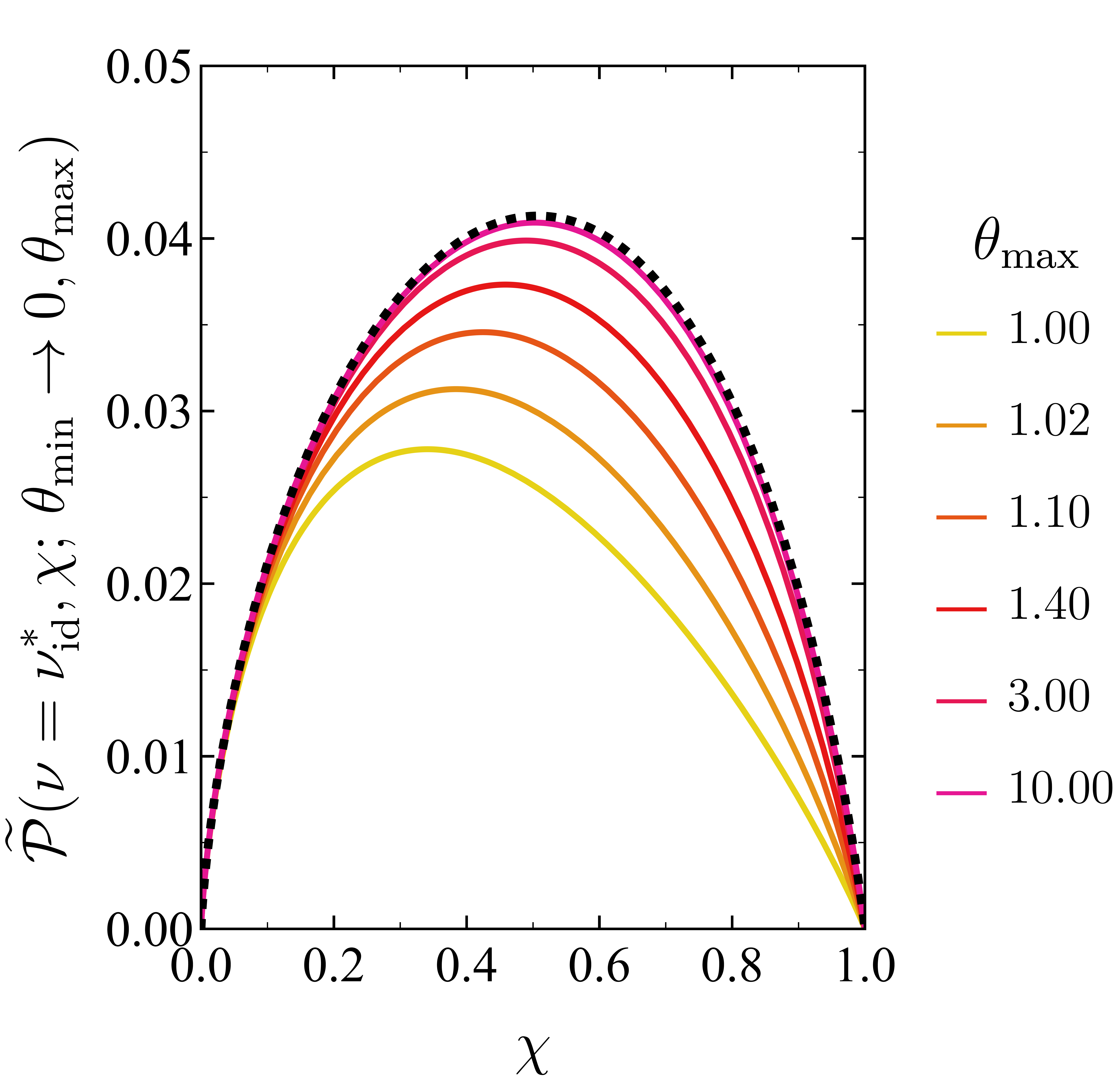}
	\end{subfigure}
	\caption{Optimal power $\widetilde{\mathcal{P}}$ as a function the compression ratio $\chi$, for fixed temperature ratio $\nu$ and different values of the temperature bounds $(\theta_{\min},\theta_{\max})$. In the four panels, we consider two values of $\nu$, $\nu=0.5$ (top) and $\nu=\nu^*_{\text{id}}$ (bottom), together with $\theta_{\max}\to +\infty$ (left) and $\theta_{\min}\to 0^{+}$ (right). In this way, we have the ideal upper (lower) bound of the temperature in the left (right) panels, whereas several different values of the other, non-ideal, temperature bound are considered. The optimal power corresponding to the ideal limit of both bounds is also displayed in all the panels (dotted black line), which is reached when the non-ideal temperature bound approaches its ideal value.}
	\label{fig:tempMmVariables_P_vs_chi}
\end{figure}

Intuitively, one expects the most beneficial scenario to correspond to the ideal bounds studied in Sec.~\ref{sec:opt-ideal-case}, i.e. $(\theta_{\min}\to0^+,\theta_{\max}\to\infty)$. To illustrate how more realistic bounds for the thermal control impinge on the optimal power, we present the behaviour of $\widetilde{\mathcal{P}}$ as a function of $\chi$ in Fig.~\ref{fig:tempMmVariables_P_vs_chi}, for multiple non-ideal values of one of the bounds whereas the other one remains ideal. We do so for two meaningful values of the temperature ratio: $\nu=0.5$ and $\nu=\nu^*_{\text{id}}$, where $\nu^*_{\text{id}}$ denotes the temperature ratio yielding the overall maximum power in the ideal limit $\theta_{\min}\to0^+$, $\theta_{\max}\to\infty$.\footnote{In the previous section, this parameter was simply written as $\nu^*$.} Indeed, the optimal power for the ideal bounds is always above the corresponding value for more realistic limits in the thermal control.

In what follows, the optimisation procedure and the notation employed is completely analogous to that in Sec.~\ref{sec:opt-ideal-case} for ideal temperature bounds. First, we address the optimisation of $\widetilde{\mathcal{P}}\left(\nu,\chi;\,\theta_{\min},\theta_{\max}\right)$, given by Eq.~\eqref{power4parameters}, over the compression ratio $\chi$. Here, the optimal compression ratio $\chi^{*}$ depends not only on $\nu$ but also on the temperature bounds, i.e. $\chi^*=\chi^*\left(\nu;\,\theta_{\min},\theta_{\max}\right)$. Therefore, we have the associated maximum power for fixed $\nu$
\begin{equation}\label{eq:P*-with-bounds}
	\widetilde{\mathcal{P}}^*\left(\nu;\,\theta_{\min},\theta_{\max}\right)\equiv \max_{\chi\in(0,1)}	\widetilde{\mathcal{P}}\left(\nu,\chi;\,\theta_{\min},\theta_{\max}\right)=\widetilde{\mathcal{P}}\left(\nu,\chi^*;\,\theta_{\min},\theta_{\max}\right)
\end{equation}
is also a function of the temperature ratio and the bounds in the thermal control. 
Again, similarly to the analysis in Sec.~\ref{sec:opt-ideal-case}, it is relevant to carry out a further optimisation over the temperature ratio $\nu$, i.e. to look for the value $\nu^{*}$ that maximises Eq.~\eqref{eq:P*-with-bounds}
\begin{equation}
  \widetilde{\mathcal{P}}^{**}\left(\theta_{\min},\theta_{\max}\right)\equiv
  \max_{\substack{
		\chi\in(0,1)\\
		\nu\in(\theta_{\min},1)
              }}\widetilde{\mathcal{P}}\left(\nu,\chi;\,\theta_{\min},\theta_{\max}\right)=
            \max_{\nu\in(\theta_{\min},1)}
            \widetilde{\mathcal{P}}^*\left(\nu;\,\theta_{\min},\theta_{\max}\right).
\end{equation}
The maximum of $\widetilde{\mathcal{P}}^{*}$ is reached at $\nu^{*}$, which now depends on the temperature bounds, $\nu^*=\nu^*\left(\theta_{\min},\theta_{\max}\right)$. Once more, similarly to the notation in Sec.~\ref{sec:opt-ideal-case}, we introduce $\chi^{**}\left(\theta_{\min},\theta_{\max}\right)\equiv\chi^*\left(\nu^*;\,\theta_{\min},\theta_{\max}\right)$. In this way, one has
\begin{equation}
  \widetilde{\mathcal{P}}^{**}\left(\theta_{\min},\theta_{\max}\right)=
  \widetilde{\mathcal{P}}^*\left(\nu^{*};\,\theta_{\min},\theta_{\max}\right)=
  \widetilde{\mathcal{P}}\left(\nu^{*},\chi^{**};\,\theta_{\min},\theta_{\max}\right).
\end{equation}
Finally, and consistently, the efficiency at overall maximum power is $\widetilde{\eta}^{**}\left(\theta_{\min},\theta_{\max}\right)\equiv\widetilde{\eta}\left(\nu^*,\chi^{**};\,\theta_{\min},\theta_{\max}\right)$.

\begin{figure}
	\begin{subfigure}{.48\textwidth}
		\raggedright
		\includegraphics[width=0.95\linewidth]{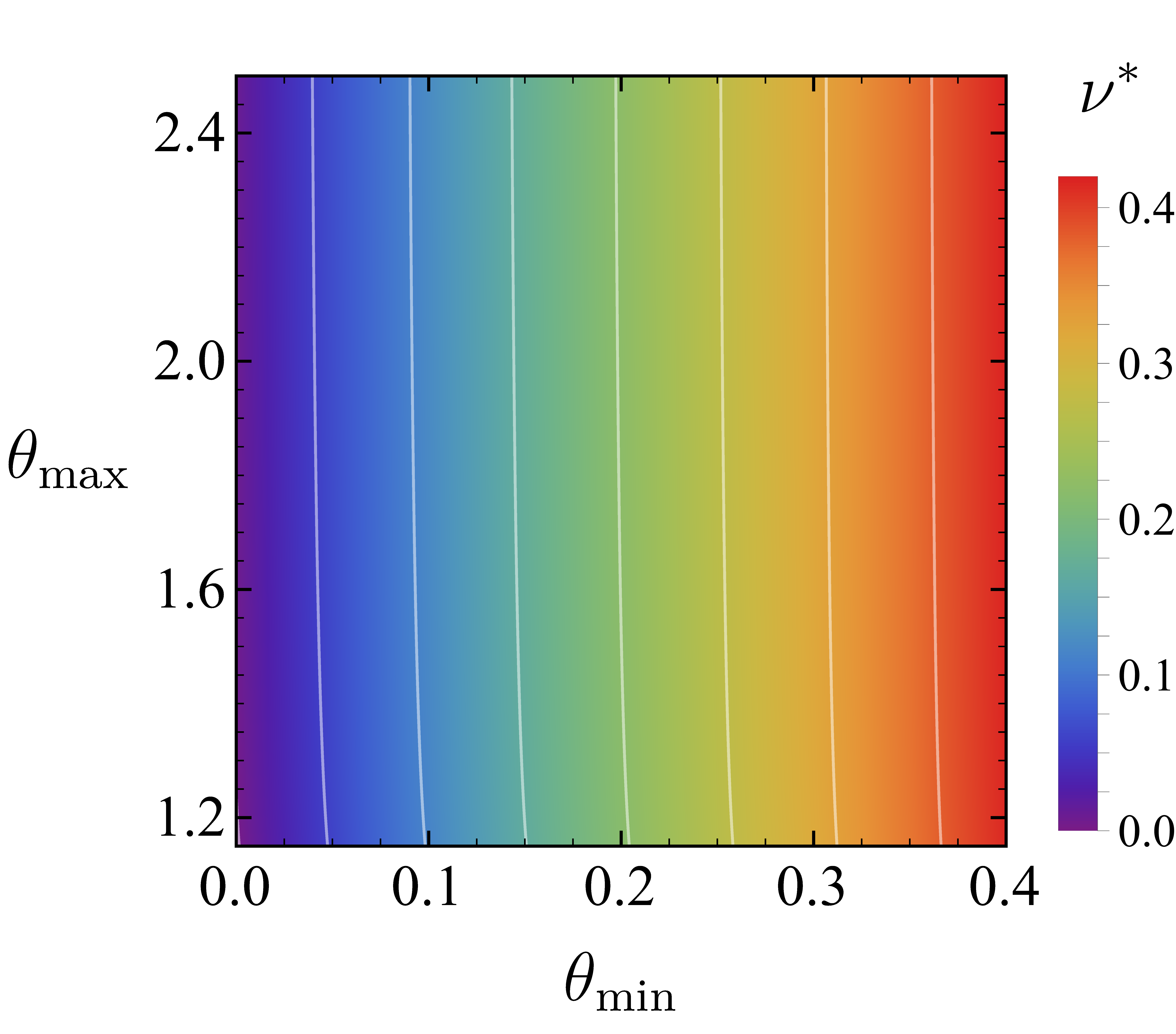}
	\end{subfigure}%
	\begin{subfigure}{.48\textwidth}
		\raggedleft
		\includegraphics[width=0.95\linewidth]{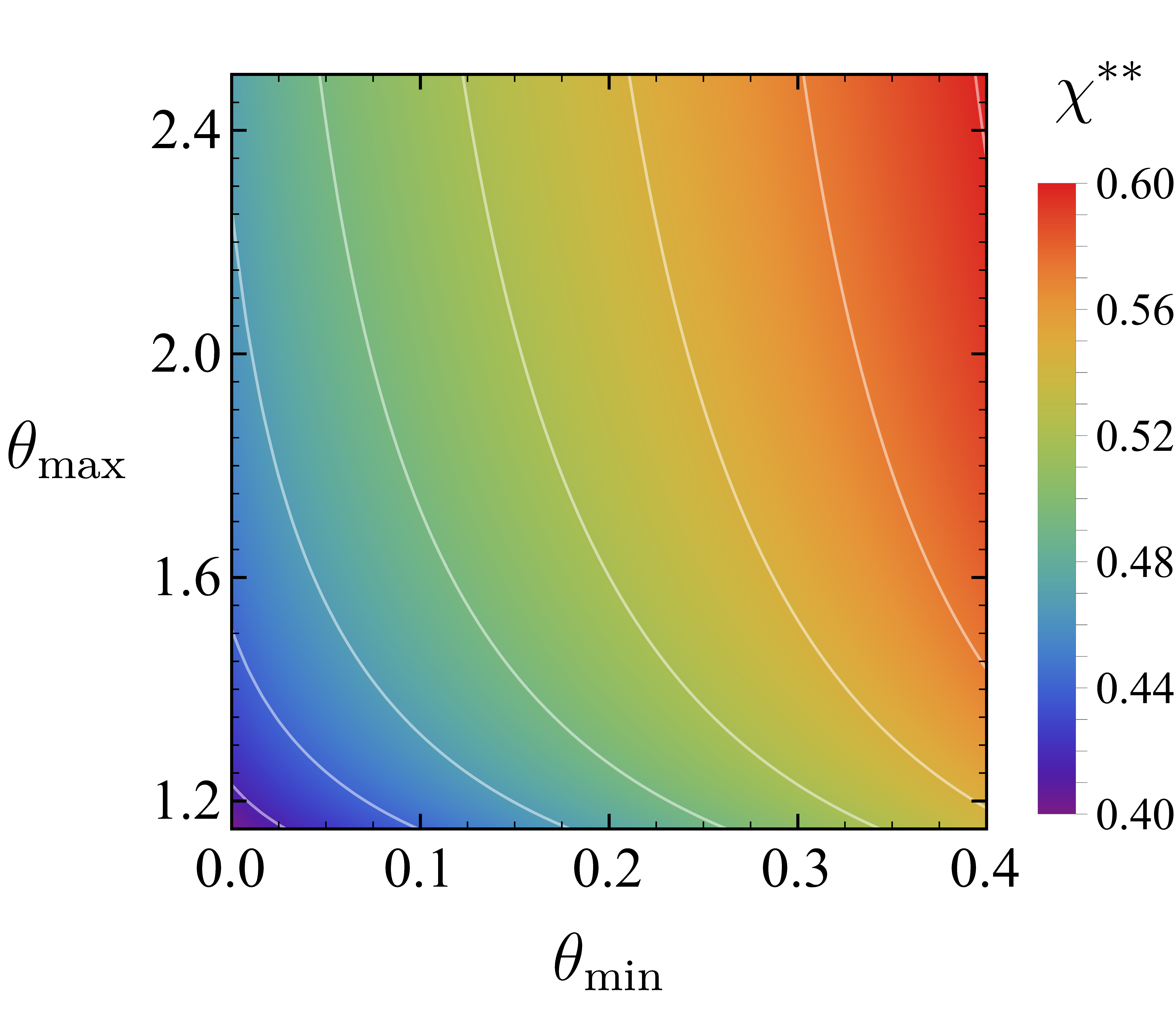}
	\end{subfigure}
		\begin{subfigure}{.48\textwidth}
		\raggedleft
		\includegraphics[width=0.95\linewidth]{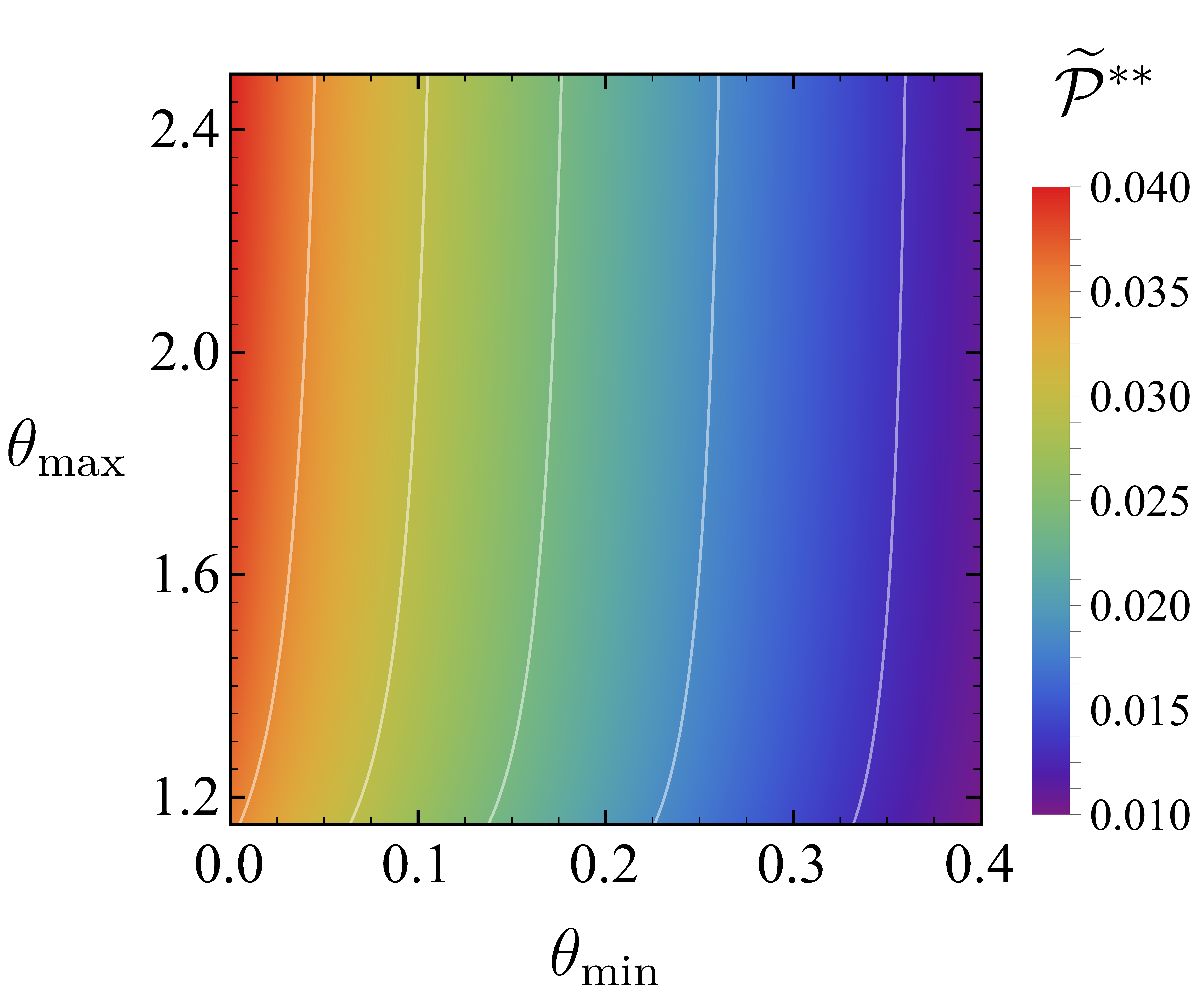}
	\end{subfigure}
	\begin{subfigure}{.48\textwidth}
		\raggedleft
		\includegraphics[width=0.95\linewidth]{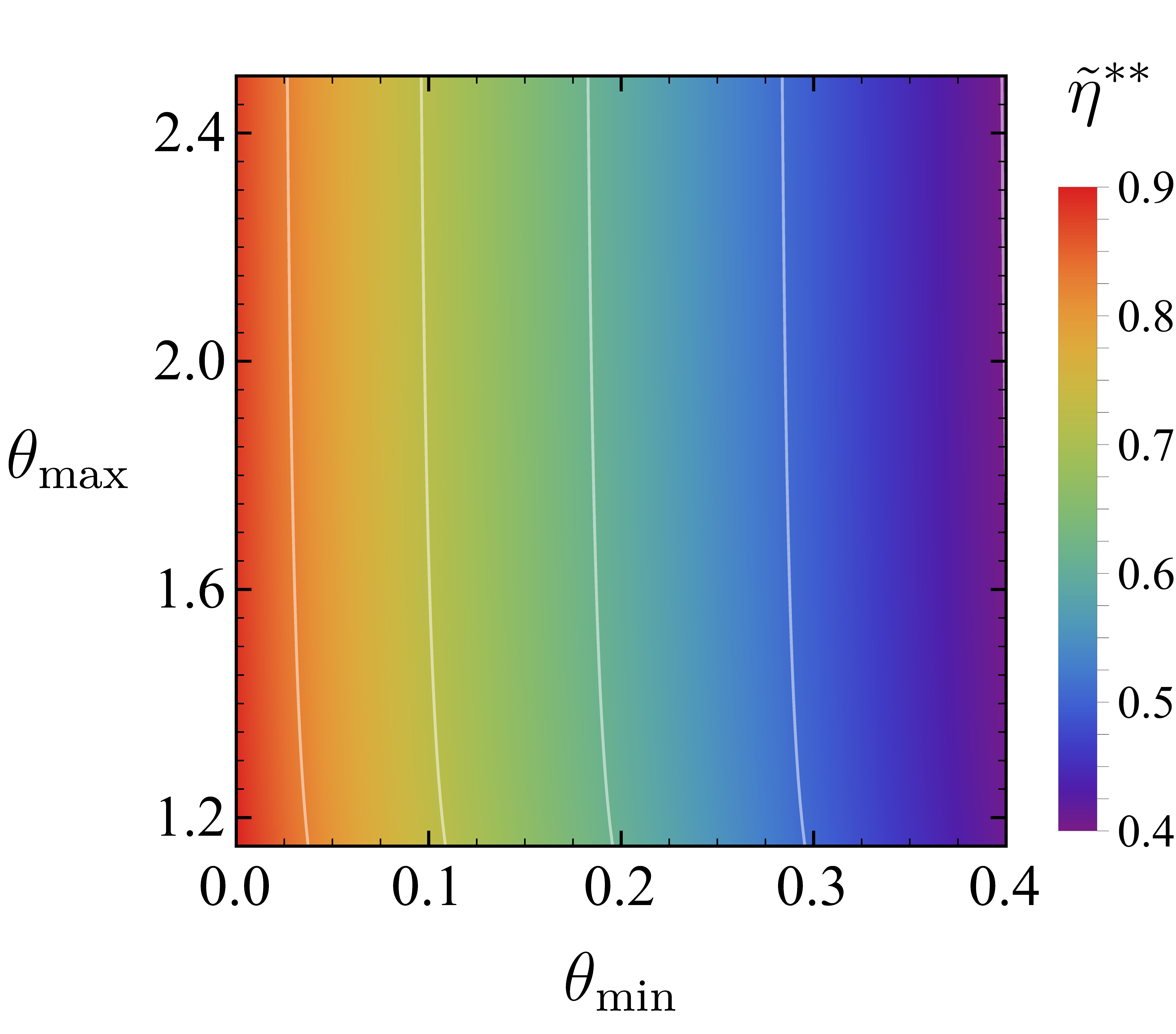}
	\end{subfigure}
	\caption{Density plots of the temperature and compression ratios (top panels) yielding optimal power (bottom left panel), and the corresponding efficiency (bottom right panel) in the $\left(\theta_{\min},\theta_{\max}\right)$ plane. We have considered the region defined by intervals $0<\theta_{\min}<0.4$, $1.15<\theta_{\max}<2.50$.}
	\label{fig:tempMmVariablesDPs}
\end{figure}

Figure~~\ref{fig:tempMmVariablesDPs} shows the overall optimal values $\nu^{*}$, $\chi^{**}$, $\widetilde{\mathcal{P}}^{**}$ and the corresponding efficiency at maximum power $\widetilde{\eta}^{**}$ as a function of $(\theta_{\min},\theta_{\max})$. Specifically, we have considered the paralellogram defined by the inequalities $0<\theta_{\min}<0.4$ and $1.15<\theta_{\max}<2.50$, inside which we have numerically solved the optimisation problem.  The maximum power $\widetilde{\mathcal{P}}^{**}$ is reduced as tightest temperature bounds are imposed, i.e. as $\theta_{\min}$ increases or $\theta_{\max}$ decreases, as expected on a physical basis. 

\section{Discussion} \label{ch:concl}
In the present work, we have theoretically designed and optimised an irreversible Stirling-like heat engine, modelled by an overdamped Brownian particle trapped in a harmonic potential. This model describes accurately the actual dynamics of trapped colloidal particles \cite{ciliberto_experiments_2017}, supporting the realisation of experiments as those in Ref.~\cite{blickle_realization_2012}. It is worth remarking that the analysis carried out here is exact and has been done for arbitrary irreversible processes in all the branches of the cycle. No assumptions, such as the low-dissipation limit \cite{Contreras_efficiency_23}, have been introduced to simplify the problem.

In the optimisation procedure, the existence of bounds on the temperature of the baths has been taken into account. Specifically, we have considered that the bath temperature $\theta$ verifies $\theta_{\min}\le \theta\le \theta_{\max}$. Our Stirling-lke engine has been analysed both in the ideal case of the loosest possible physical bounds, $(\theta_{\min}\to 0^{+},\theta_{\max}\to+\infty)$, and in the realistic case of $\theta$ being restricted to a finite interval. 

Remarkably, the values obtained for the optimal value of the power in the ideal case beat those found in both the original experimental realization of a the Brownian Stirling-like heat engine \cite{blickle_realization_2012}---by a factor 100---and the theoretical proposal for a Brownian Carnot-like heat engine \cite{plata_building_2020}---by a factor 10. Regarding the comparison with the latter, the Stirling-like cycle's outperfoming Carnot's makes physical sense: although the temperatures of the isotherms are the same in both cycles, the use of more extreme temperatures during the isochoric branches, $\theta_{\min}$ for the cooling one and  $\theta_{\max} )$ for the heating one,  can be understood as having better thermal resources in the Stirling engine---thus leading to a better performance in terms of power.

The hypothesis of overdamped dynamics is expected to break down as the inertial effects become more relevant, e.g. by increasing the stiffness of the trap. However, since the stiffness scale is freely chosen at the beginning through the choice of the  operating point A---fixing the reference units, see Table \ref{table:operating_points}---the underdamped regime can be avoided with a careful choice thereof.

Interestingly from a practical point of view,  the efficiency at maximum power for the optimal Stirling-like engine is  quite high, $\eta \simeq 0.8$, for ideal temperature bounds. Moreover, we have shown that the efficiency at maximum power always surpasses the Curzon-Alhborn value, for arbitrary operating points and temperature bounds. It is important to keep in mind that this is not a violation of any physical law, since the Curzon-Alhborn value represents a bound just for specific conditions. When considering systems with strong coupling between work and heat fluxes \cite{van_den_broeck_eff} in the linear response at maximum power, the Curzon-Ahlborn efficiency turns out to be an upper bound for the efficiency at maximum power in the linear response regime. A clear identification of the microscopic ingredients leading to the Curzon-Ahlborn efficiency as an upper bound for Brownian heat engines has recently been carried out \cite{Quan_CA_22}.

Possible future research work, stemming from the results derived in this paper, is discussed in the following.  Present-day experimental techniques make it possible to control both the stiffness of the trap---by using optical tweezers---and the temperature of the bath---by applying a random electric field with controlled amplitude. As discussed above, experimental realisations of micrometre-sized stochastic Stirling engines have already been carried out \cite{blickle_realization_2012}. Therefore, it would be interesting to implement the proposed optimal Stirling-like engine in an actual experiment. On another note, investigating exactly---beyond the low dissipation regime---the irreversible analogue of other classical thermodynamic cycles, such as Otto's, Diesel's or Ericsson's, may lead to interesting results, both on the theoretical and experimental contexts.

\subsection*{Acknowledgements}

A.P. and C.A.P. acknowledge financial support from Grants PID2021-122588NB-I00, funded by {MCIN/AEI/10.13039/} {501100011033/} and by ``ERDF A way of making Europe'', and ProyExcel\_00796, funded by Junta de Andalucía's PAIDI 2020 programme. C.A.P acknowledges the funding received from European Union’s Horizon Europe–Marie Skłodowska-Curie 2021 program through the Postdoctoral Fellowship with Reference 101065902 (ORION). I.P.R. acknowledges the financial support received from the Spanish Ministry of Education through the Collaboration Grant programme 2022/2023.

\subsection*{Data availability}

The Mathematica notebooks employed for generating the data and figures that support the findings of this study are openly available in the~\href{https://github.com/fine-group-us/Max-power-Stirling-Brownian-heat-engine}{GitHub page} of University of Sevilla's FINE research group.

\bibliographystyle{apsrev4-2}
\bibliography{Biblio_corregida}

\begin{thebibliography}{59}%
\makeatletter
\providecommand \@ifxundefined [1]{%
 \@ifx{#1\undefined}
}%
\providecommand \@ifnum [1]{%
 \ifnum #1\expandafter \@firstoftwo
 \else \expandafter \@secondoftwo
 \fi
}%
\providecommand \@ifx [1]{%
 \ifx #1\expandafter \@firstoftwo
 \else \expandafter \@secondoftwo
 \fi
}%
\providecommand \natexlab [1]{#1}%
\providecommand \enquote  [1]{``#1''}%
\providecommand \bibnamefont  [1]{#1}%
\providecommand \bibfnamefont [1]{#1}%
\providecommand \citenamefont [1]{#1}%
\providecommand \href@noop [0]{\@secondoftwo}%
\providecommand \href [0]{\begingroup \@sanitize@url \@href}%
\providecommand \@href[1]{\@@startlink{#1}\@@href}%
\providecommand \@@href[1]{\endgroup#1\@@endlink}%
\providecommand \@sanitize@url [0]{\catcode `\\12\catcode `\$12\catcode
  `\&12\catcode `\#12\catcode `\^12\catcode `\_12\catcode `\%12\relax}%
\providecommand \@@startlink[1]{}%
\providecommand \@@endlink[0]{}%
\providecommand \url  [0]{\begingroup\@sanitize@url \@url }%
\providecommand \@url [1]{\endgroup\@href {#1}{\urlprefix }}%
\providecommand \urlprefix  [0]{URL }%
\providecommand \Eprint [0]{\href }%
\providecommand \doibase [0]{https://doi.org/}%
\providecommand \selectlanguage [0]{\@gobble}%
\providecommand \bibinfo  [0]{\@secondoftwo}%
\providecommand \bibfield  [0]{\@secondoftwo}%
\providecommand \translation [1]{[#1]}%
\providecommand \BibitemOpen [0]{}%
\providecommand \bibitemStop [0]{}%
\providecommand \bibitemNoStop [0]{.\EOS\space}%
\providecommand \EOS [0]{\spacefactor3000\relax}%
\providecommand \BibitemShut  [1]{\csname bibitem#1\endcsname}%
\let\auto@bib@innerbib\@empty
%</preamble>
\bibitem [{\citenamefont {Sekimoto}(2010)}]{sekimoto_book}%
  \BibitemOpen
  \bibfield  {author} {\bibinfo {author} {\bibfnamefont {K.}~\bibnamefont
  {Sekimoto}},\ }\href@noop {} {\emph {\bibinfo {title} {Stochastic
  {Energetics}}}}\ (\bibinfo  {publisher} {Springer},\ \bibinfo {year}
  {2010})\BibitemShut {NoStop}%
\bibitem [{\citenamefont {Shiraishi}(2023)}]{shiraishi_book}%
  \BibitemOpen
  \bibfield  {author} {\bibinfo {author} {\bibfnamefont {N.}~\bibnamefont
  {Shiraishi}},\ }\href {https://books.google.es/books?id=I6O9EAAAQBAJ} {\emph
  {\bibinfo {title} {An Introduction to Stochastic Thermodynamics: From Basic
  to Advanced}}},\ Fundamental Theories of Physics\ (\bibinfo  {publisher}
  {Springer Nature Singapore},\ \bibinfo {year} {2023})\BibitemShut {NoStop}%
\bibitem [{\citenamefont {Seifert}(2012)}]{seifert_stochastic_2012}%
  \BibitemOpen
  \bibfield  {author} {\bibinfo {author} {\bibfnamefont {U.}~\bibnamefont
  {Seifert}},\ }\href {https://doi.org/10.1088/0034-4885/75/12/126001}
  {\bibfield  {journal} {\bibinfo  {journal} {Reports on Progress in Physics}\
  }\textbf {\bibinfo {volume} {75}},\ \bibinfo {pages} {126001} (\bibinfo
  {year} {2012})}\BibitemShut {NoStop}%
\bibitem [{\citenamefont {Van~Kampen}(1992)}]{vankampen}%
  \BibitemOpen
  \bibfield  {author} {\bibinfo {author} {\bibfnamefont {N.~G.}\ \bibnamefont
  {Van~Kampen}},\ }\href@noop {} {\emph {\bibinfo {title} {Stochastic Processes
  in Physics and Chemistry}}}\ (\bibinfo  {publisher} {Elsevier},\ \bibinfo
  {year} {1992})\BibitemShut {NoStop}%
\bibitem [{\citenamefont {Ciliberto}(2017)}]{ciliberto_experiments_2017}%
  \BibitemOpen
  \bibfield  {author} {\bibinfo {author} {\bibfnamefont {S.}~\bibnamefont
  {Ciliberto}},\ }\href {https://doi.org/10.1103/PhysRevX.7.021051} {\bibfield
  {journal} {\bibinfo  {journal} {Physical Review X}\ }\textbf {\bibinfo
  {volume} {7}},\ \bibinfo {pages} {021051} (\bibinfo {year}
  {2017})}\BibitemShut {NoStop}%
\bibitem [{\citenamefont {Thomson}(1879)}]{noauthor_sorting_1879}%
  \BibitemOpen
  \bibfield  {author} {\bibinfo {author} {\bibfnamefont {W.}~\bibnamefont
  {Thomson}},\ }\href {https://doi.org/10.1038/020126a0} {\bibfield  {journal}
  {\bibinfo  {journal} {Nature}\ }\textbf {\bibinfo {volume} {20}},\ \bibinfo
  {pages} {126} (\bibinfo {year} {1879})}\BibitemShut {NoStop}%
\bibitem [{\citenamefont {Mart\'{\i}nez}\ \emph {et~al.}(2017)\citenamefont
  {Mart\'{\i}nez}, \citenamefont {Rold\'an}, \citenamefont {Dinis},\ and\
  \citenamefont {Rica}}]{martinez_colloidal_2017}%
  \BibitemOpen
  \bibfield  {author} {\bibinfo {author} {\bibfnamefont {I.~A.}\ \bibnamefont
  {Mart\'{\i}nez}}, \bibinfo {author} {\bibfnamefont {E.}~\bibnamefont
  {Rold\'an}}, \bibinfo {author} {\bibfnamefont {L.}~\bibnamefont {Dinis}},\
  and\ \bibinfo {author} {\bibfnamefont {R.~A.}\ \bibnamefont {Rica}},\ }\href
  {https://doi.org/10.1039/C6SM00923A} {\bibfield  {journal} {\bibinfo
  {journal} {Soft Matter}\ }\textbf {\bibinfo {volume} {13}},\ \bibinfo {pages}
  {22} (\bibinfo {year} {2017})}\BibitemShut {NoStop}%
\bibitem [{\citenamefont {Magnasco}(1993)}]{Magnasco_ratchet_93}%
  \BibitemOpen
  \bibfield  {author} {\bibinfo {author} {\bibfnamefont {M.~O.}\ \bibnamefont
  {Magnasco}},\ }\href {https://doi.org/10.1103/PhysRevLett.71.1477} {\bibfield
   {journal} {\bibinfo  {journal} {Phys. Rev. Lett.}\ }\textbf {\bibinfo
  {volume} {71}},\ \bibinfo {pages} {1477} (\bibinfo {year}
  {1993})}\BibitemShut {NoStop}%
\bibitem [{\citenamefont {Reimann}\ \emph {et~al.}(1996)\citenamefont
  {Reimann}, \citenamefont {Bartussek}, \citenamefont {Häußler},\ and\
  \citenamefont {Hänggi}}]{Reimann_motor_96}%
  \BibitemOpen
  \bibfield  {author} {\bibinfo {author} {\bibfnamefont {P.}~\bibnamefont
  {Reimann}}, \bibinfo {author} {\bibfnamefont {R.}~\bibnamefont {Bartussek}},
  \bibinfo {author} {\bibfnamefont {R.}~\bibnamefont {Häußler}},\ and\
  \bibinfo {author} {\bibfnamefont {P.}~\bibnamefont {Hänggi}},\ }\href
  {https://doi.org/https://doi.org/10.1016/0375-9601(96)00222-8} {\bibfield
  {journal} {\bibinfo  {journal} {Physics Letters A}\ }\textbf {\bibinfo
  {volume} {215}},\ \bibinfo {pages} {26} (\bibinfo {year} {1996})}\BibitemShut
  {NoStop}%
\bibitem [{\citenamefont {J\"ulicher}\ \emph {et~al.}(1997)\citenamefont
  {J\"ulicher}, \citenamefont {Ajdari},\ and\ \citenamefont
  {Prost}}]{Julicher_motor_97}%
  \BibitemOpen
  \bibfield  {author} {\bibinfo {author} {\bibfnamefont {F.}~\bibnamefont
  {J\"ulicher}}, \bibinfo {author} {\bibfnamefont {A.}~\bibnamefont {Ajdari}},\
  and\ \bibinfo {author} {\bibfnamefont {J.}~\bibnamefont {Prost}},\ }\href
  {https://doi.org/10.1103/RevModPhys.69.1269} {\bibfield  {journal} {\bibinfo
  {journal} {Rev. Mod. Phys.}\ }\textbf {\bibinfo {volume} {69}},\ \bibinfo
  {pages} {1269} (\bibinfo {year} {1997})}\BibitemShut {NoStop}%
\bibitem [{\citenamefont {Reimann}(2002)}]{Reimann_motor_02}%
  \BibitemOpen
  \bibfield  {author} {\bibinfo {author} {\bibfnamefont {P.}~\bibnamefont
  {Reimann}},\ }\href
  {https://doi.org/https://doi.org/10.1016/S0370-1573(01)00081-3} {\bibfield
  {journal} {\bibinfo  {journal} {Physics Reports}\ }\textbf {\bibinfo {volume}
  {361}},\ \bibinfo {pages} {57} (\bibinfo {year} {2002})}\BibitemShut
  {NoStop}%
\bibitem [{\citenamefont {Hänggi}\ \emph {et~al.}(2005)\citenamefont
  {Hänggi}, \citenamefont {Marchesoni},\ and\ \citenamefont
  {Nori}}]{Hanggi_motor_05}%
  \BibitemOpen
  \bibfield  {author} {\bibinfo {author} {\bibfnamefont {P.}~\bibnamefont
  {Hänggi}}, \bibinfo {author} {\bibfnamefont {F.}~\bibnamefont
  {Marchesoni}},\ and\ \bibinfo {author} {\bibfnamefont {F.}~\bibnamefont
  {Nori}},\ }\href {https://doi.org/https://doi.org/10.1002/andp.200551701-304}
  {\bibfield  {journal} {\bibinfo  {journal} {Annalen der Physik}\ }\textbf
  {\bibinfo {volume} {517}},\ \bibinfo {pages} {51} (\bibinfo {year} {2005})},\
  \Eprint
  {https://arxiv.org/abs/https://onlinelibrary.wiley.com/doi/pdf/10.1002/andp.200551701-304}
  {https://onlinelibrary.wiley.com/doi/pdf/10.1002/andp.200551701-304}
  \BibitemShut {NoStop}%
\bibitem [{\citenamefont {Horowitz}\ and\ \citenamefont
  {Parrondo}(2011)}]{horowitz11}%
  \BibitemOpen
  \bibfield  {author} {\bibinfo {author} {\bibfnamefont {J.}~\bibnamefont
  {Horowitz}}\ and\ \bibinfo {author} {\bibfnamefont {J.~M.}\ \bibnamefont
  {Parrondo}},\ }\href {https://doi.org/10.1038/nphys2184} {\bibfield
  {journal} {\bibinfo  {journal} {Nature Physics}\ }\textbf {\bibinfo {volume}
  {8}},\ \bibinfo {pages} {108} (\bibinfo {year} {2011})}\BibitemShut {NoStop}%
\bibitem [{\citenamefont {Blickle}\ and\ \citenamefont
  {Bechinger}(2012)}]{blickle_realization_2012}%
  \BibitemOpen
  \bibfield  {author} {\bibinfo {author} {\bibfnamefont {V.}~\bibnamefont
  {Blickle}}\ and\ \bibinfo {author} {\bibfnamefont {C.}~\bibnamefont
  {Bechinger}},\ }\href {https://doi.org/10.1038/nphys2163} {\bibfield
  {journal} {\bibinfo  {journal} {Nature Physics}\ }\textbf {\bibinfo {volume}
  {8}},\ \bibinfo {pages} {143} (\bibinfo {year} {2012})}\BibitemShut {NoStop}%
\bibitem [{\citenamefont {Holubec}(2014)}]{Holubec_logpot_14}%
  \BibitemOpen
  \bibfield  {author} {\bibinfo {author} {\bibfnamefont {V.}~\bibnamefont
  {Holubec}},\ }\href {https://doi.org/10.1088/1742-5468/2014/05/P05022}
  {\bibfield  {journal} {\bibinfo  {journal} {Journal of Statistical Mechanics:
  Theory and Experiment}\ }\textbf {\bibinfo {volume} {2014}},\ \bibinfo
  {pages} {P05022} (\bibinfo {year} {2014})}\BibitemShut {NoStop}%
\bibitem [{\citenamefont {Rana}\ \emph {et~al.}(2014)\citenamefont {Rana},
  \citenamefont {Pal}, \citenamefont {Saha},\ and\ \citenamefont
  {Jayannavar}}]{Rana_Single_14}%
  \BibitemOpen
  \bibfield  {author} {\bibinfo {author} {\bibfnamefont {S.}~\bibnamefont
  {Rana}}, \bibinfo {author} {\bibfnamefont {P.~S.}\ \bibnamefont {Pal}},
  \bibinfo {author} {\bibfnamefont {A.}~\bibnamefont {Saha}},\ and\ \bibinfo
  {author} {\bibfnamefont {A.~M.}\ \bibnamefont {Jayannavar}},\ }\href
  {https://doi.org/10.1103/PhysRevE.90.042146} {\bibfield  {journal} {\bibinfo
  {journal} {Phys. Rev. E}\ }\textbf {\bibinfo {volume} {90}},\ \bibinfo
  {pages} {042146} (\bibinfo {year} {2014})}\BibitemShut {NoStop}%
\bibitem [{\citenamefont {Tu}(2014)}]{Tu_inertial_14}%
  \BibitemOpen
  \bibfield  {author} {\bibinfo {author} {\bibfnamefont {Z.~C.}\ \bibnamefont
  {Tu}},\ }\href {https://doi.org/10.1103/PhysRevE.89.052148} {\bibfield
  {journal} {\bibinfo  {journal} {Phys. Rev. E}\ }\textbf {\bibinfo {volume}
  {89}},\ \bibinfo {pages} {052148} (\bibinfo {year} {2014})}\BibitemShut
  {NoStop}%
\bibitem [{\citenamefont {Mart\'{\i}nez}\ \emph {et~al.}(2015)\citenamefont
  {Mart\'{\i}nez}, \citenamefont {Rold\'an}, \citenamefont {Dinis},
  \citenamefont {Petrov}, \citenamefont {Parrondo},\ and\ \citenamefont
  {Rica}}]{martinez15BrownianCarnotEngine}%
  \BibitemOpen
  \bibfield  {author} {\bibinfo {author} {\bibfnamefont {I.}~\bibnamefont
  {Mart\'{\i}nez}}, \bibinfo {author} {\bibfnamefont {E.}~\bibnamefont
  {Rold\'an}}, \bibinfo {author} {\bibfnamefont {L.}~\bibnamefont {Dinis}},
  \bibinfo {author} {\bibfnamefont {D.}~\bibnamefont {Petrov}}, \bibinfo
  {author} {\bibfnamefont {J.~M.}\ \bibnamefont {Parrondo}},\ and\ \bibinfo
  {author} {\bibfnamefont {R.}~\bibnamefont {Rica}},\ }\href
  {https://doi.org/10.1038/nphys3518} {\bibfield  {journal} {\bibinfo
  {journal} {Nature Physics}\ }\textbf {\bibinfo {volume} {12}},\ \bibinfo
  {pages} {67} (\bibinfo {year} {2015})}\BibitemShut {NoStop}%
\bibitem [{\citenamefont {Bauer}\ \emph {et~al.}(2016)\citenamefont {Bauer},
  \citenamefont {Brandner},\ and\ \citenamefont {Seifert}}]{Bauer_limited_16}%
  \BibitemOpen
  \bibfield  {author} {\bibinfo {author} {\bibfnamefont {M.}~\bibnamefont
  {Bauer}}, \bibinfo {author} {\bibfnamefont {K.}~\bibnamefont {Brandner}},\
  and\ \bibinfo {author} {\bibfnamefont {U.}~\bibnamefont {Seifert}},\ }\href
  {https://doi.org/10.1103/PhysRevE.93.042112} {\bibfield  {journal} {\bibinfo
  {journal} {Phys. Rev. E}\ }\textbf {\bibinfo {volume} {93}},\ \bibinfo
  {pages} {042112} (\bibinfo {year} {2016})}\BibitemShut {NoStop}%
\bibitem [{\citenamefont {Dechant}\ \emph {et~al.}(2017)\citenamefont
  {Dechant}, \citenamefont {Kiesel},\ and\ \citenamefont
  {Lutz}}]{Dechant_underdamped_17}%
  \BibitemOpen
  \bibfield  {author} {\bibinfo {author} {\bibfnamefont {A.}~\bibnamefont
  {Dechant}}, \bibinfo {author} {\bibfnamefont {N.}~\bibnamefont {Kiesel}},\
  and\ \bibinfo {author} {\bibfnamefont {E.}~\bibnamefont {Lutz}},\ }\href
  {https://doi.org/10.1209/0295-5075/119/50003} {\bibfield  {journal} {\bibinfo
   {journal} {Europhysics Letters}\ }\textbf {\bibinfo {volume} {119}},\
  \bibinfo {pages} {50003} (\bibinfo {year} {2017})}\BibitemShut {NoStop}%
\bibitem [{\citenamefont {Plata}\ \emph
  {et~al.}(2020{\natexlab{a}})\citenamefont {Plata}, \citenamefont
  {Guéry-Odelin}, \citenamefont {Trizac},\ and\ \citenamefont
  {Prados}}]{plata_building_2020}%
  \BibitemOpen
  \bibfield  {author} {\bibinfo {author} {\bibfnamefont {C.~A.}\ \bibnamefont
  {Plata}}, \bibinfo {author} {\bibfnamefont {D.}~\bibnamefont
  {Guéry-Odelin}}, \bibinfo {author} {\bibfnamefont {E.}~\bibnamefont
  {Trizac}},\ and\ \bibinfo {author} {\bibfnamefont {A.}~\bibnamefont
  {Prados}},\ }\href {https://doi.org/10.1088/1742-5468/abb0e1} {\bibfield
  {journal} {\bibinfo  {journal} {Journal of Statistical Mechanics: Theory and
  Experiment}\ ,\ \bibinfo {pages} {093207}} (\bibinfo {year}
  {2020}{\natexlab{a}})}\BibitemShut {NoStop}%
\bibitem [{\citenamefont {Tu}(2021)}]{Tu_abstract_21}%
  \BibitemOpen
  \bibfield  {author} {\bibinfo {author} {\bibfnamefont {Z.-C.}\ \bibnamefont
  {Tu}},\ }\href {https://doi.org/10.1007/s11467-020-1029-6} {\bibfield
  {journal} {\bibinfo  {journal} {Frontiers of Physics}\ }\textbf {\bibinfo
  {volume} {16}},\ \bibinfo {pages} {33202} (\bibinfo {year}
  {2021})}\BibitemShut {NoStop}%
\bibitem [{\citenamefont {Apertet}\ \emph
  {et~al.}(2012{\natexlab{a}})\citenamefont {Apertet}, \citenamefont
  {Ouerdane}, \citenamefont {Goupil},\ and\ \citenamefont
  {Lecoeur}}]{Apertet_thermoelectric_12}%
  \BibitemOpen
  \bibfield  {author} {\bibinfo {author} {\bibfnamefont {Y.}~\bibnamefont
  {Apertet}}, \bibinfo {author} {\bibfnamefont {H.}~\bibnamefont {Ouerdane}},
  \bibinfo {author} {\bibfnamefont {C.}~\bibnamefont {Goupil}},\ and\ \bibinfo
  {author} {\bibfnamefont {P.}~\bibnamefont {Lecoeur}},\ }\href
  {https://doi.org/10.1103/PhysRevE.85.031116} {\bibfield  {journal} {\bibinfo
  {journal} {Phys. Rev. E}\ }\textbf {\bibinfo {volume} {85}},\ \bibinfo
  {pages} {031116} (\bibinfo {year} {2012}{\natexlab{a}})}\BibitemShut
  {NoStop}%
\bibitem [{\citenamefont {Apertet}\ \emph
  {et~al.}(2012{\natexlab{b}})\citenamefont {Apertet}, \citenamefont
  {Ouerdane}, \citenamefont {Goupil},\ and\ \citenamefont
  {Lecoeur}}]{Apertet_thermoelectric2_12}%
  \BibitemOpen
  \bibfield  {author} {\bibinfo {author} {\bibfnamefont {Y.}~\bibnamefont
  {Apertet}}, \bibinfo {author} {\bibfnamefont {H.}~\bibnamefont {Ouerdane}},
  \bibinfo {author} {\bibfnamefont {C.}~\bibnamefont {Goupil}},\ and\ \bibinfo
  {author} {\bibfnamefont {P.}~\bibnamefont {Lecoeur}},\ }\href
  {https://doi.org/10.1103/PhysRevE.85.041144} {\bibfield  {journal} {\bibinfo
  {journal} {Phys. Rev. E}\ }\textbf {\bibinfo {volume} {85}},\ \bibinfo
  {pages} {041144} (\bibinfo {year} {2012}{\natexlab{b}})}\BibitemShut
  {NoStop}%
\bibitem [{\citenamefont {Ouerdane}\ \emph {et~al.}(2015)\citenamefont
  {Ouerdane}, \citenamefont {Apertet}, \citenamefont {Goupil},\ and\
  \citenamefont {Lecoeur}}]{Ouerdane_thermoelectric_15}%
  \BibitemOpen
  \bibfield  {author} {\bibinfo {author} {\bibfnamefont {H.}~\bibnamefont
  {Ouerdane}}, \bibinfo {author} {\bibfnamefont {Y.}~\bibnamefont {Apertet}},
  \bibinfo {author} {\bibfnamefont {C.}~\bibnamefont {Goupil}},\ and\ \bibinfo
  {author} {\bibfnamefont {P.}~\bibnamefont {Lecoeur}},\ }\href
  {https://doi.org/10.1140/epjst/e2015-02431-x} {\bibfield  {journal} {\bibinfo
   {journal} {European Physical Journal: Special Topics}\ }\textbf {\bibinfo
  {volume} {224}},\ \bibinfo {pages} {839 – 862} (\bibinfo {year} {2015})},\
  \bibinfo {note} {cited by: 31}\BibitemShut {NoStop}%
\bibitem [{\citenamefont {Hua}\ and\ \citenamefont
  {Guo}(2024)}]{Hua_thermoelectric_24}%
  \BibitemOpen
  \bibfield  {author} {\bibinfo {author} {\bibfnamefont {Y.}~\bibnamefont
  {Hua}}\ and\ \bibinfo {author} {\bibfnamefont {Z.-Y.}\ \bibnamefont {Guo}},\
  }\href {https://doi.org/10.1103/PhysRevE.109.024130} {\bibfield  {journal}
  {\bibinfo  {journal} {Phys. Rev. E}\ }\textbf {\bibinfo {volume} {109}},\
  \bibinfo {pages} {024130} (\bibinfo {year} {2024})}\BibitemShut {NoStop}%
\bibitem [{\citenamefont {Gingrich}\ \emph {et~al.}(2014)\citenamefont
  {Gingrich}, \citenamefont {Rotskoff}, \citenamefont {Vaikuntanathan},\ and\
  \citenamefont {Geissler}}]{Gingrich_LargeDeviation_14}%
  \BibitemOpen
  \bibfield  {author} {\bibinfo {author} {\bibfnamefont {T.~R.}\ \bibnamefont
  {Gingrich}}, \bibinfo {author} {\bibfnamefont {G.~M.}\ \bibnamefont
  {Rotskoff}}, \bibinfo {author} {\bibfnamefont {S.}~\bibnamefont
  {Vaikuntanathan}},\ and\ \bibinfo {author} {\bibfnamefont {P.~L.}\
  \bibnamefont {Geissler}},\ }\href
  {https://doi.org/10.1088/1367-2630/16/10/102003} {\bibfield  {journal}
  {\bibinfo  {journal} {New Journal of Physics}\ }\textbf {\bibinfo {volume}
  {16}},\ \bibinfo {pages} {102003} (\bibinfo {year} {2014})}\BibitemShut
  {NoStop}%
\bibitem [{\citenamefont {Krishnamurthy}\ \emph {et~al.}(2016)\citenamefont
  {Krishnamurthy}, \citenamefont {Ghosh}, \citenamefont {Chatterji},
  \citenamefont {Ganapathy},\ and\ \citenamefont
  {Sood}}]{Krishnamurthy_active_16}%
  \BibitemOpen
  \bibfield  {author} {\bibinfo {author} {\bibfnamefont {S.}~\bibnamefont
  {Krishnamurthy}}, \bibinfo {author} {\bibfnamefont {S.}~\bibnamefont
  {Ghosh}}, \bibinfo {author} {\bibfnamefont {D.}~\bibnamefont {Chatterji}},
  \bibinfo {author} {\bibfnamefont {R.}~\bibnamefont {Ganapathy}},\ and\
  \bibinfo {author} {\bibfnamefont {A.}~\bibnamefont {Sood}},\ }\href
  {https://doi.org/10.1038/nphys3870} {\bibfield  {journal} {\bibinfo
  {journal} {Nature Physics}\ }\textbf {\bibinfo {volume} {12}},\ \bibinfo
  {pages} {1134 – 1138} (\bibinfo {year} {2016})},\ \bibinfo {note} {cited
  by: 166}\BibitemShut {NoStop}%
\bibitem [{\citenamefont {Kumari}\ \emph {et~al.}(2020)\citenamefont {Kumari},
  \citenamefont {Pal}, \citenamefont {Saha},\ and\ \citenamefont
  {Lahiri}}]{Kumari_active_20}%
  \BibitemOpen
  \bibfield  {author} {\bibinfo {author} {\bibfnamefont {A.}~\bibnamefont
  {Kumari}}, \bibinfo {author} {\bibfnamefont {P.~S.}\ \bibnamefont {Pal}},
  \bibinfo {author} {\bibfnamefont {A.}~\bibnamefont {Saha}},\ and\ \bibinfo
  {author} {\bibfnamefont {S.}~\bibnamefont {Lahiri}},\ }\href
  {https://doi.org/10.1103/PhysRevE.101.032109} {\bibfield  {journal} {\bibinfo
   {journal} {Phys. Rev. E}\ }\textbf {\bibinfo {volume} {101}},\ \bibinfo
  {pages} {032109} (\bibinfo {year} {2020})}\BibitemShut {NoStop}%
\bibitem [{\citenamefont {Shiraishi}\ \emph {et~al.}(2016)\citenamefont
  {Shiraishi}, \citenamefont {Saito},\ and\ \citenamefont
  {Tasaki}}]{Shiraishi_trade-off_16}%
  \BibitemOpen
  \bibfield  {author} {\bibinfo {author} {\bibfnamefont {N.}~\bibnamefont
  {Shiraishi}}, \bibinfo {author} {\bibfnamefont {K.}~\bibnamefont {Saito}},\
  and\ \bibinfo {author} {\bibfnamefont {H.}~\bibnamefont {Tasaki}},\ }\href
  {https://doi.org/10.1103/PhysRevLett.117.190601} {\bibfield  {journal}
  {\bibinfo  {journal} {Phys. Rev. Lett.}\ }\textbf {\bibinfo {volume} {117}},\
  \bibinfo {pages} {190601} (\bibinfo {year} {2016})}\BibitemShut {NoStop}%
\bibitem [{\citenamefont {Pietzonka}\ and\ \citenamefont
  {Seifert}(2018)}]{Pietzonka_trade-off_18}%
  \BibitemOpen
  \bibfield  {author} {\bibinfo {author} {\bibfnamefont {P.}~\bibnamefont
  {Pietzonka}}\ and\ \bibinfo {author} {\bibfnamefont {U.}~\bibnamefont
  {Seifert}},\ }\href {https://doi.org/10.1103/PhysRevLett.120.190602}
  {\bibfield  {journal} {\bibinfo  {journal} {Phys. Rev. Lett.}\ }\textbf
  {\bibinfo {volume} {120}},\ \bibinfo {pages} {190602} (\bibinfo {year}
  {2018})}\BibitemShut {NoStop}%
\bibitem [{\citenamefont {Curzon}\ and\ \citenamefont
  {Ahlborn}(1975)}]{curzon_efficiency_1975}%
  \BibitemOpen
  \bibfield  {author} {\bibinfo {author} {\bibfnamefont {F.~L.}\ \bibnamefont
  {Curzon}}\ and\ \bibinfo {author} {\bibfnamefont {B.}~\bibnamefont
  {Ahlborn}},\ }\href {https://doi.org/10.1119/1.10023} {\bibfield  {journal}
  {\bibinfo  {journal} {American Journal of Physics}\ }\textbf {\bibinfo
  {volume} {43}},\ \bibinfo {pages} {22} (\bibinfo {year} {1975})}\BibitemShut
  {NoStop}%
\bibitem [{\citenamefont {Van~den Broeck}(2005)}]{van_den_broeck_eff}%
  \BibitemOpen
  \bibfield  {author} {\bibinfo {author} {\bibfnamefont {C.}~\bibnamefont
  {Van~den Broeck}},\ }\href {https://doi.org/10.1103/PhysRevLett.95.190602}
  {\bibfield  {journal} {\bibinfo  {journal} {Physical Review Letters}\
  }\textbf {\bibinfo {volume} {95}},\ \bibinfo {pages} {190602} (\bibinfo
  {year} {2005})}\BibitemShut {NoStop}%
\bibitem [{\citenamefont {Esposito}\ \emph {et~al.}(2009)\citenamefont
  {Esposito}, \citenamefont {Lindenberg},\ and\ \citenamefont {Van~den
  Broeck}}]{esposito_universality_2009}%
  \BibitemOpen
  \bibfield  {author} {\bibinfo {author} {\bibfnamefont {M.}~\bibnamefont
  {Esposito}}, \bibinfo {author} {\bibfnamefont {K.}~\bibnamefont
  {Lindenberg}},\ and\ \bibinfo {author} {\bibfnamefont {C.}~\bibnamefont
  {Van~den Broeck}},\ }\href {https://doi.org/10.1103/PhysRevLett.102.130602}
  {\bibfield  {journal} {\bibinfo  {journal} {Physical Review Letters}\
  }\textbf {\bibinfo {volume} {102}},\ \bibinfo {pages} {130602} (\bibinfo
  {year} {2009})}\BibitemShut {NoStop}%
\bibitem [{\citenamefont {Esposito}\ \emph {et~al.}(2010)\citenamefont
  {Esposito}, \citenamefont {Kawai}, \citenamefont {Lindenberg},\ and\
  \citenamefont {Van~den Broeck}}]{esposito_efficiency_2010}%
  \BibitemOpen
  \bibfield  {author} {\bibinfo {author} {\bibfnamefont {M.}~\bibnamefont
  {Esposito}}, \bibinfo {author} {\bibfnamefont {R.}~\bibnamefont {Kawai}},
  \bibinfo {author} {\bibfnamefont {K.}~\bibnamefont {Lindenberg}},\ and\
  \bibinfo {author} {\bibfnamefont {C.}~\bibnamefont {Van~den Broeck}},\ }\href
  {https://doi.org/10.1103/PhysRevLett.105.150603} {\bibfield  {journal}
  {\bibinfo  {journal} {Physical Review Letters}\ }\textbf {\bibinfo {volume}
  {105}},\ \bibinfo {pages} {150603} (\bibinfo {year} {2010})}\BibitemShut
  {NoStop}%
\bibitem [{\citenamefont {Wang}\ and\ \citenamefont
  {Tu}(2012)}]{Wang_efficiency_12}%
  \BibitemOpen
  \bibfield  {author} {\bibinfo {author} {\bibfnamefont {Y.}~\bibnamefont
  {Wang}}\ and\ \bibinfo {author} {\bibfnamefont {Z.~C.}\ \bibnamefont {Tu}},\
  }\href {https://doi.org/10.1103/PhysRevE.85.011127} {\bibfield  {journal}
  {\bibinfo  {journal} {Phys. Rev. E}\ }\textbf {\bibinfo {volume} {85}},\
  \bibinfo {pages} {011127} (\bibinfo {year} {2012})}\BibitemShut {NoStop}%
\bibitem [{\citenamefont {Tlili}(2012)}]{tlili_finite_2012}%
  \BibitemOpen
  \bibfield  {author} {\bibinfo {author} {\bibfnamefont {I.}~\bibnamefont
  {Tlili}},\ }\href {https://doi.org/10.1016/j.rser.2012.01.022} {\bibfield
  {journal} {\bibinfo  {journal} {Renewable and Sustainable Energy Reviews}\
  }\textbf {\bibinfo {volume} {16}},\ \bibinfo {pages} {2234} (\bibinfo {year}
  {2012})}\BibitemShut {NoStop}%
\bibitem [{\citenamefont {Gonzalez-Ayala}\ \emph {et~al.}(2017)\citenamefont
  {Gonzalez-Ayala}, \citenamefont {Calvo~Hern\'andez},\ and\ \citenamefont
  {Roco}}]{Gonzalez-Ayala_trade-off_17}%
  \BibitemOpen
  \bibfield  {author} {\bibinfo {author} {\bibfnamefont {J.}~\bibnamefont
  {Gonzalez-Ayala}}, \bibinfo {author} {\bibfnamefont {A.}~\bibnamefont
  {Calvo~Hern\'andez}},\ and\ \bibinfo {author} {\bibfnamefont {J.~M.~M.}\
  \bibnamefont {Roco}},\ }\href {https://doi.org/10.1103/PhysRevE.95.022131}
  {\bibfield  {journal} {\bibinfo  {journal} {Phys. Rev. E}\ }\textbf {\bibinfo
  {volume} {95}},\ \bibinfo {pages} {022131} (\bibinfo {year}
  {2017})}\BibitemShut {NoStop}%
\bibitem [{\citenamefont {Frim}\ and\ \citenamefont
  {DeWeese}(2022{\natexlab{a}})}]{Frim_Geometric_bound_22}%
  \BibitemOpen
  \bibfield  {author} {\bibinfo {author} {\bibfnamefont {A.~G.}\ \bibnamefont
  {Frim}}\ and\ \bibinfo {author} {\bibfnamefont {M.~R.}\ \bibnamefont
  {DeWeese}},\ }\href {https://doi.org/10.1103/PhysRevLett.128.230601}
  {\bibfield  {journal} {\bibinfo  {journal} {Phys. Rev. Lett.}\ }\textbf
  {\bibinfo {volume} {128}},\ \bibinfo {pages} {230601} (\bibinfo {year}
  {2022}{\natexlab{a}})}\BibitemShut {NoStop}%
\bibitem [{\citenamefont {Frim}\ and\ \citenamefont
  {DeWeese}(2022{\natexlab{b}})}]{Frim_Optimal_22}%
  \BibitemOpen
  \bibfield  {author} {\bibinfo {author} {\bibfnamefont {A.~G.}\ \bibnamefont
  {Frim}}\ and\ \bibinfo {author} {\bibfnamefont {M.~R.}\ \bibnamefont
  {DeWeese}},\ }\href {https://doi.org/10.1103/PhysRevE.105.L052103} {\bibfield
   {journal} {\bibinfo  {journal} {Phys. Rev. E}\ }\textbf {\bibinfo {volume}
  {105}},\ \bibinfo {pages} {L052103} (\bibinfo {year}
  {2022}{\natexlab{b}})}\BibitemShut {NoStop}%
\bibitem [{\citenamefont {Contreras-Vergara}\ \emph {et~al.}(2023)\citenamefont
  {Contreras-Vergara}, \citenamefont {S\'anchez-Salas}, \citenamefont
  {Valencia-Ortega},\ and\ \citenamefont
  {Jim\'enez-Aquino}}]{Contreras_efficiency_23}%
  \BibitemOpen
  \bibfield  {author} {\bibinfo {author} {\bibfnamefont {O.}~\bibnamefont
  {Contreras-Vergara}}, \bibinfo {author} {\bibfnamefont {N.}~\bibnamefont
  {S\'anchez-Salas}}, \bibinfo {author} {\bibfnamefont {G.}~\bibnamefont
  {Valencia-Ortega}},\ and\ \bibinfo {author} {\bibfnamefont {J.~I.}\
  \bibnamefont {Jim\'enez-Aquino}},\ }\href
  {https://doi.org/10.1103/PhysRevE.108.014123} {\bibfield  {journal} {\bibinfo
   {journal} {Phys. Rev. E}\ }\textbf {\bibinfo {volume} {108}},\ \bibinfo
  {pages} {014123} (\bibinfo {year} {2023})}\BibitemShut {NoStop}%
\bibitem [{\citenamefont {Schmiedl}\ and\ \citenamefont
  {Seifert}(2008)}]{schmiedl_efficiency_2008}%
  \BibitemOpen
  \bibfield  {author} {\bibinfo {author} {\bibfnamefont {T.}~\bibnamefont
  {Schmiedl}}\ and\ \bibinfo {author} {\bibfnamefont {U.}~\bibnamefont
  {Seifert}},\ }\href {https://doi.org/10.1209/0295-5075/81/20003} {\bibfield
  {journal} {\bibinfo  {journal} {EPL (Europhysics Letters)}\ }\textbf
  {\bibinfo {volume} {81}},\ \bibinfo {pages} {20003} (\bibinfo {year}
  {2008})}\BibitemShut {NoStop}%
\bibitem [{\citenamefont {Nakamura}\ \emph {et~al.}(2020)\citenamefont
  {Nakamura}, \citenamefont {Matrasulov},\ and\ \citenamefont
  {Izumida}}]{nakamura_fast_2020}%
  \BibitemOpen
  \bibfield  {author} {\bibinfo {author} {\bibfnamefont {K.}~\bibnamefont
  {Nakamura}}, \bibinfo {author} {\bibfnamefont {J.}~\bibnamefont
  {Matrasulov}},\ and\ \bibinfo {author} {\bibfnamefont {Y.}~\bibnamefont
  {Izumida}},\ }\href {https://doi.org/10.1103/PhysRevE.102.012129} {\bibfield
  {journal} {\bibinfo  {journal} {Physical Review E}\ }\textbf {\bibinfo
  {volume} {102}},\ \bibinfo {pages} {012129} (\bibinfo {year}
  {2020})}\BibitemShut {NoStop}%
\bibitem [{\citenamefont {Schmiedl}\ and\ \citenamefont
  {Seifert}(2007)}]{schmiedl_optimal_2007}%
  \BibitemOpen
  \bibfield  {author} {\bibinfo {author} {\bibfnamefont {T.}~\bibnamefont
  {Schmiedl}}\ and\ \bibinfo {author} {\bibfnamefont {U.}~\bibnamefont
  {Seifert}},\ }\href {https://doi.org/10.1103/PhysRevLett.98.108301}
  {\bibfield  {journal} {\bibinfo  {journal} {Physical Review Letters}\
  }\textbf {\bibinfo {volume} {98}},\ \bibinfo {pages} {108301} (\bibinfo
  {year} {2007})}\BibitemShut {NoStop}%
\bibitem [{\citenamefont {Plata}\ \emph
  {et~al.}(2020{\natexlab{b}})\citenamefont {Plata}, \citenamefont
  {Guéry-Odelin}, \citenamefont {Trizac},\ and\ \citenamefont
  {Prados}}]{plata_finite-time_2020}%
  \BibitemOpen
  \bibfield  {author} {\bibinfo {author} {\bibfnamefont {C.~A.}\ \bibnamefont
  {Plata}}, \bibinfo {author} {\bibfnamefont {D.}~\bibnamefont
  {Guéry-Odelin}}, \bibinfo {author} {\bibfnamefont {E.}~\bibnamefont
  {Trizac}},\ and\ \bibinfo {author} {\bibfnamefont {A.}~\bibnamefont
  {Prados}},\ }\href {https://doi.org/10.1103/PhysRevE.101.032129} {\bibfield
  {journal} {\bibinfo  {journal} {Physical Review E}\ }\textbf {\bibinfo
  {volume} {101}},\ \bibinfo {pages} {032129} (\bibinfo {year}
  {2020}{\natexlab{b}})}\BibitemShut {NoStop}%
\bibitem [{\citenamefont {Patrón}\ \emph {et~al.}(2022)\citenamefont
  {Patrón}, \citenamefont {Prados},\ and\ \citenamefont
  {Plata}}]{patron_thermal_2022}%
  \BibitemOpen
  \bibfield  {author} {\bibinfo {author} {\bibfnamefont {A.}~\bibnamefont
  {Patrón}}, \bibinfo {author} {\bibfnamefont {A.}~\bibnamefont {Prados}},\
  and\ \bibinfo {author} {\bibfnamefont {C.~A.}\ \bibnamefont {Plata}},\ }\href
  {https://doi.org/10.1140/epjp/s13360-022-03150-3} {\bibfield  {journal}
  {\bibinfo  {journal} {The European Physical Journal Plus}\ }\textbf {\bibinfo
  {volume} {137}},\ \bibinfo {pages} {1011} (\bibinfo {year}
  {2022})}\BibitemShut {NoStop}%
\bibitem [{\citenamefont {Pontryagin}(1987)}]{pontryagin}%
  \BibitemOpen
  \bibfield  {author} {\bibinfo {author} {\bibfnamefont {L.~S.}\ \bibnamefont
  {Pontryagin}},\ }\href@noop {} {\emph {\bibinfo {title} {Mathematical Theory
  of Optimal Processes}}}\ (\bibinfo  {publisher} {CRC Press},\ \bibinfo {year}
  {1987})\BibitemShut {NoStop}%
\bibitem [{\citenamefont {Liberzon}(2012)}]{liberzon}%
  \BibitemOpen
  \bibfield  {author} {\bibinfo {author} {\bibfnamefont {D.}~\bibnamefont
  {Liberzon}},\ }\href {http://www.jstor.org/stable/j.ctvcm4g0s} {\emph
  {\bibinfo {title} {Calculus of Variations and Optimal Control Theory: A
  Concise Introduction}}}\ (\bibinfo  {publisher} {Princeton University
  Press},\ \bibinfo {year} {2012})\BibitemShut {NoStop}%
\bibitem [{\citenamefont {Martínez}\ \emph {et~al.}(2016)\citenamefont
  {Martínez}, \citenamefont {Petrosyan}, \citenamefont {Guéry-Odelin},
  \citenamefont {Trizac},\ and\ \citenamefont
  {Ciliberto}}]{martinez_engineered_2016}%
  \BibitemOpen
  \bibfield  {author} {\bibinfo {author} {\bibfnamefont {I.~A.}\ \bibnamefont
  {Martínez}}, \bibinfo {author} {\bibfnamefont {A.}~\bibnamefont
  {Petrosyan}}, \bibinfo {author} {\bibfnamefont {D.}~\bibnamefont
  {Guéry-Odelin}}, \bibinfo {author} {\bibfnamefont {E.}~\bibnamefont
  {Trizac}},\ and\ \bibinfo {author} {\bibfnamefont {S.}~\bibnamefont
  {Ciliberto}},\ }\href {https://doi.org/10.1038/nphys3758} {\bibfield
  {journal} {\bibinfo  {journal} {Nature Physics}\ }\textbf {\bibinfo {volume}
  {12}},\ \bibinfo {pages} {843} (\bibinfo {year} {2016})}\BibitemShut
  {NoStop}%
\bibitem [{\citenamefont {Guéry-Odelin}\ \emph {et~al.}(2023)\citenamefont
  {Guéry-Odelin}, \citenamefont {Jarzynski}, \citenamefont {Plata},
  \citenamefont {Prados},\ and\ \citenamefont
  {Trizac}}]{guery-odelin_driving_2023}%
  \BibitemOpen
  \bibfield  {author} {\bibinfo {author} {\bibfnamefont {D.}~\bibnamefont
  {Guéry-Odelin}}, \bibinfo {author} {\bibfnamefont {C.}~\bibnamefont
  {Jarzynski}}, \bibinfo {author} {\bibfnamefont {C.~A.}\ \bibnamefont
  {Plata}}, \bibinfo {author} {\bibfnamefont {A.}~\bibnamefont {Prados}},\ and\
  \bibinfo {author} {\bibfnamefont {E.}~\bibnamefont {Trizac}},\ }\href
  {https://doi.org/10.1088/1361-6633/acacad} {\bibfield  {journal} {\bibinfo
  {journal} {Reports on Progress in Physics}\ }\textbf {\bibinfo {volume}
  {86}},\ \bibinfo {pages} {035902} (\bibinfo {year} {2023})}\BibitemShut
  {NoStop}%
\bibitem [{\citenamefont {Mart\'{\i}nez}\ \emph {et~al.}(2013)\citenamefont
  {Mart\'{\i}nez}, \citenamefont {Rold\'an}, \citenamefont {Parrondo},\ and\
  \citenamefont {Petrov}}]{martinez_effective_13}%
  \BibitemOpen
  \bibfield  {author} {\bibinfo {author} {\bibfnamefont {I.~A.}\ \bibnamefont
  {Mart\'{\i}nez}}, \bibinfo {author} {\bibfnamefont {E.}~\bibnamefont
  {Rold\'an}}, \bibinfo {author} {\bibfnamefont {J.~M.~R.}\ \bibnamefont
  {Parrondo}},\ and\ \bibinfo {author} {\bibfnamefont {D.}~\bibnamefont
  {Petrov}},\ }\href {https://doi.org/10.1103/PhysRevE.87.032159} {\bibfield
  {journal} {\bibinfo  {journal} {Phys. Rev. E}\ }\textbf {\bibinfo {volume}
  {87}},\ \bibinfo {pages} {032159} (\bibinfo {year} {2013})}\BibitemShut
  {NoStop}%
\bibitem [{\citenamefont {Callen}(1985)}]{callen}%
  \BibitemOpen
  \bibfield  {author} {\bibinfo {author} {\bibfnamefont {H.~B.}\ \bibnamefont
  {Callen}},\ }\href@noop {} {\emph {\bibinfo {title} {Thermodynamics and an
  Introduction to Thermostatistics}}}\ (\bibinfo  {publisher} {Wiley},\
  \bibinfo {year} {1985})\BibitemShut {NoStop}%
\bibitem [{\citenamefont {Plata}\ \emph {et~al.}(2019)\citenamefont {Plata},
  \citenamefont {Guéry-Odelin}, \citenamefont {Trizac},\ and\ \citenamefont
  {Prados}}]{plata_optimal_2019}%
  \BibitemOpen
  \bibfield  {author} {\bibinfo {author} {\bibfnamefont {C.~A.}\ \bibnamefont
  {Plata}}, \bibinfo {author} {\bibfnamefont {D.}~\bibnamefont
  {Guéry-Odelin}}, \bibinfo {author} {\bibfnamefont {E.}~\bibnamefont
  {Trizac}},\ and\ \bibinfo {author} {\bibfnamefont {A.}~\bibnamefont
  {Prados}},\ }\href {https://doi.org/10.1103/PhysRevE.99.012140} {\bibfield
  {journal} {\bibinfo  {journal} {Physical Review E}\ }\textbf {\bibinfo
  {volume} {99}},\ \bibinfo {pages} {012140} (\bibinfo {year}
  {2019})}\BibitemShut {NoStop}%
\bibitem [{\citenamefont {Band}\ \emph {et~al.}(1982)\citenamefont {Band},
  \citenamefont {Kafri},\ and\ \citenamefont {Salamon}}]{band82}%
  \BibitemOpen
  \bibfield  {author} {\bibinfo {author} {\bibfnamefont {Y.}~\bibnamefont
  {Band}}, \bibinfo {author} {\bibfnamefont {O.}~\bibnamefont {Kafri}},\ and\
  \bibinfo {author} {\bibfnamefont {P.}~\bibnamefont {Salamon}},\ }\href
  {https://doi.org/10.1063/1.329960} {\bibfield  {journal} {\bibinfo  {journal}
  {Journal of Applied Physics}\ }\textbf {\bibinfo {volume} {53}},\ \bibinfo
  {pages} {8 } (\bibinfo {year} {1982})}\BibitemShut {NoStop}%
\bibitem [{\citenamefont {Kaushik}\ \emph {et~al.}(2017)\citenamefont
  {Kaushik}, \citenamefont {Tyagi},\ and\ \citenamefont
  {Kumar}}]{finite_thermo_book}%
  \BibitemOpen
  \bibfield  {author} {\bibinfo {author} {\bibfnamefont {S.~C.}\ \bibnamefont
  {Kaushik}}, \bibinfo {author} {\bibfnamefont {S.~K.}\ \bibnamefont {Tyagi}},\
  and\ \bibinfo {author} {\bibfnamefont {P.}~\bibnamefont {Kumar}},\ }\bibinfo
  {title} {Finite time thermodynamic analysis of stirling and ericsson power
  cycles},\ in\ \href {https://doi.org/10.1007/978-3-319-62812-7_6} {\emph
  {\bibinfo {booktitle} {Finite Time Thermodynamics of Power and Refrigeration
  Cycles}}}\ (\bibinfo  {publisher} {Springer International Publishing},\
  \bibinfo {address} {Cham},\ \bibinfo {year} {2017})\ pp.\ \bibinfo {pages}
  {115--148}\BibitemShut {NoStop}%
\bibitem [{Note1()}]{Note1}%
  \BibitemOpen
  \bibinfo {note} {For the estimation of the delivered power in the experiment
  reported in Ref.~\cite {blickle_realization_2012}, we have taken into account
  that the typical drag coefficient of a micrometre-sized bead in water
  solution is around $\lambda = 10^{-8}\protect \tmspace +\thickmuskip
  {.2777em} \protect \text {kg}\protect \tmspace +\thickmuskip {.2777em}
  \protect \text {s}^{-1}$, whereas the stiffness in that experiment was around
  $k=10^{-6} \protect \tmspace +\thickmuskip {.2777em}\protect \text {kg}
  \protect \tmspace +\thickmuskip {.2777em} \protect \text {s}^{-2}$ and the
  cold and hot bath temperature were $22\protect \tmspace +\thickmuskip
  {.2777em}^\circ $C and $86\protect \tmspace +\thickmuskip {.2777em}^\circ
  $C.}\BibitemShut {Stop}%
\bibitem [{Note2()}]{Note2}%
  \BibitemOpen
  \bibinfo {note} {In the limit $\chi \to 0^{+}$, the minimum time over the
  isochores diverge, whereas in the limit $\chi \to 1^{-}$, the maximum work
  vanishes.}\BibitemShut {Stop}%
\bibitem [{Note3()}]{Note3}%
  \BibitemOpen
  \bibinfo {note} {In the previous section, this parameter was simply written
  as $\nu ^*$.}\BibitemShut {Stop}%
\bibitem [{\citenamefont {Chen}\ \emph {et~al.}(2022)\citenamefont {Chen},
  \citenamefont {Chen}, \citenamefont {Fei},\ and\ \citenamefont
  {Quan}}]{Quan_CA_22}%
  \BibitemOpen
  \bibfield  {author} {\bibinfo {author} {\bibfnamefont {Y.~H.}\ \bibnamefont
  {Chen}}, \bibinfo {author} {\bibfnamefont {J.-F.}\ \bibnamefont {Chen}},
  \bibinfo {author} {\bibfnamefont {Z.}~\bibnamefont {Fei}},\ and\ \bibinfo
  {author} {\bibfnamefont {H.~T.}\ \bibnamefont {Quan}},\ }\href
  {https://doi.org/10.1103/PhysRevE.106.024105} {\bibfield  {journal} {\bibinfo
   {journal} {Phys. Rev. E}\ }\textbf {\bibinfo {volume} {106}},\ \bibinfo
  {pages} {024105} (\bibinfo {year} {2022})}\BibitemShut {NoStop}%
\end{thebibliography}%
\end{document}